\documentstyle[iopconf1]{article}
\input psfig.sty
\def\nn{\noindent}
\def\ie{{\it i.e.}}
\def\eg{{\it e.g.}}

\def\etal{{\it et al.}}

\def\epem{\ifmmode e^+e^-\else $e^+e^-$\fi}
\def\to{\rightarrow}

\def\mpl{\ifmmode \overline M_{Pl}\else $\bar M_{Pl}$\fi}

\begin{document}

\rightline{\vbox{\halign{&#\hfil\cr
SLAC-PUB-8238\cr
October 1999\cr}}}
\vspace{0.8in}

\title{{Indirect Collider Tests for Large Extra Dimensions}
\footnote{To appear in the {\it Proceedings of Beyond the Desert99:
Accelerator, Non-Accelerator and Space Approaches}, Castle Ringberg, 
Tegernsee, Germany, June 6-12 1999}
}

\author{Thomas G Rizzo\footnote{E-mail:
rizzo@slacvx.slac.stanford.edu. 
Work supported by the Department of Energy, 
Contract DE-AC03-76SF00515
}}

\affil{Stanford Linear Accelerator Center, 
Stanford CA 94309, USA}

\beginabstract
We review the capability of colliders to detect the virtual 
exchange of Kaluza-Klein 
towers of gravitons within the low scale quantum gravity scenario of 
Arkani-Hamed, Dimopoulos and Dvali.
\endabstract

\section{Introduction}

Arkani-Hamed, Dimopoulos and Dvali(ADD) recently proposed a interesting 
low scale quantum gravity scenario which offers a new slant on the hierarchy 
problem{\cite {nima}}. In its simplest form, gravity is allowed to live in 
$n$ `large' extra dimensions, \ie, `the bulk', while the Standard Model(SM) 
fields lie on a 3-D surface or brane, `the wall'. Gravity then becomes strong 
in the full $4+n$-dimensional space at a scale $M_s\sim$ a few TeV which is 
far below the conventional Planck scale, $M_{pl}\sim 10^{19}$ GeV. The scales 
$M_s$ and $M_{pl}$ are simply related via Gauss' Law:
\begin{equation}
M_{pl}^2=V_nM_s^{n+2}\,,
\end{equation}
with $V_n$ being the volume of the compactified 
extra dimensions. For $n$ extra dimensions 
of the same size, the simplest possibility, termed a symmetric 
compactification, $V_n\sim R^n$ and one finds that 
$R \sim 10^{30/n-19}$ meters assuming $M_s\sim 1$ TeV. Note that 
for separations 
between two masses less than $R$ the gravitational force law becomes 
$1/r^{2+n}$. For $n=1$, $R\sim 10^{11}$ meters and is thus obviously excluded, 
but, for $n=2$ one obtains $R \sim 0.1$~mm, which is at the edge of the 
sensitivity for existing experiments{\cite {test}}. One can imagine more 
general scenarios, termed asymmetric compactifications, where, \eg, there are 
$p$ `small' dimensions that have sizes of $\sim 1/TeV$ 
with the effective number of `large' extra dimensions being $n_{eff}=n-p$,  
where now $n=6$ or 7 as 
suggested by string theory{\cite {ln}}. Astrophysics{\cite {astro}} requires 
that $M_s>110$ TeV for $n=2$ but only $\geq $ a few TeV for $n>2$. 

The ADD scenario is based on the assumption that the metric tensor factorizes, 
\ie, that the usual 4-D components are independent of the co-ordinates of the 
other dimensions. Giving up this assumption can lead to a number of other 
interesting scenarios with completely different phenomenology{\cite {rs}}. 

The Feynman rules{\cite {pheno1}} for the ADD scenario are obtained by 
considering a linearized theory of gravity
in the bulk, decomposing it into the more familiar 4-D states and
recalling the existence of Kaluza-Klein towers for each of the conventionally
massless fields. The entire set of fields in the K-K tower couples in an
identical fashion to the particles of the SM. By considering the forms of the
$4+n$
symmetric, conserved stress-energy tensor for the various SM fields and by
remembering that such fields live only on the wall one may derive all of the
necessary couplings. An important result of
these considerations is that only the massive spin-2 K-K towers (which couple
to the 4-D conserved, symmetric stress-energy tensor, $T^{\mu\nu}$) and 
spin-0 K-K
towers (which couple proportional to the trace of $T^{\mu\nu}$) are of
phenomenological relevance as all the spin-1 fields can be shown to decouple
from the particles of the SM. For processes that involve massless fields at
at least one vertex, as will be the case below, the contributions of
the spin-0 fields can also be ignored. 
   
Outside of table top experiments and astrophysics, the two ways of probing 
this scenario are 
via the emission of KK towers of gravitons in scattering processes or through 
the exchange of KK graviton towers between SM fields{\cite {pheno1,pheno2}}. 
In the case of emission, one uses the Feynman rules to calculate the cross 
section for the production of a graviton of fixed mass and then sums over the 
full tower of states. If the mass splittings among the gravitons are small in 
comparison to the typical energy scale of the process of interest we can 
replace the summation by an integral weighted by the $n$-D density 
of states and which is cut off by the specific process kinematics. Due to the 
huge number of KK states which are integrated over the typical $M_{pl}$ 
suppression one expects from gravity vanishes and is replaced by suppression 
due to simple powers of $E/M_s$, with $E$ being the typical energy scale in 
the relevant interaction. Perhaps the 
simplest process of this kind is $e^+e^-\to G_n \gamma$, with $G_n$ 
representing the tower of gravitons that appear as missing energy in a 
detector. Fig.1 shows the typical signal cross section and background for this 
process at LEP II for different numbers of extra dimensions with $M_s=1$ TeV 
which is close to the present experimental bound{\cite {exp}} for $n=2$. 

\vspace*{-0.5cm}
\nn
\begin{figure}[htbp]
\centerline{
\psfig{figure=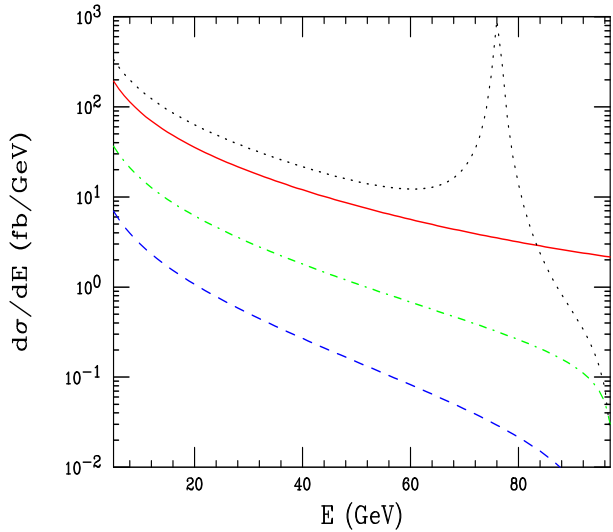,height=7.cm,width=8cm,angle=90}}
\vspace*{-0.05cm}
\caption{Cross section with $\sqrt s=195$ GeV for $e^+e^-\to G_n \gamma$ as a 
function of the photon 
energy assuming $M_s=1$ TeV with 2(3,4) extra dimensions corresponding to the 
solid(dash-dotted, dashed) curve. The dotted curve is the SM background.}
\end{figure}

\section{Graviton Exchange}

The virtual 
exchange of graviton towers either leads to modifications in SM cross 
sections and asymmetries or to new processes not allowed in the SM at the tree 
level. In the case of exchange the amplitude is proportional to the sum over 
the propagators of the entire KK tower which naively diverges when $n>1$ even 
when transformed to an integral over the density of states. This integral can 
either be regulated by a brute force cut-off, by the tension of the 
3-brane{\cite {wow}}, or through the finite extent of the SM fermion wave 
functions in the additional dimensions{\cite {schm}}. The differential cross 
sections then become relatively 
$n$ insensitive functions of the effective cut-off scale, traditionally 
taken as $M_s$, and the overall sign of the dimension-8 operator induced by 
the KK tower, $\lambda$. Thus effectively, we essentially have the replacement
\begin{equation}
{i^2\over 8\mpl^2}\sum_{n=1}^\infty{1\over s-m_n^2} \to 
{\lambda\over M_s^4}\,.
\end{equation}
where $\mpl$ is the reduced Planck scale. Note that this replacement is 
{\it universal} in that it is independent of the choice of particles in the 
initial or final state. Similar substitutions also take 
place in the $t$- and $u$-channels allowing for the straightforward 
calculation of a large number of cross sections for different processes. A 
characteristic feature in all cases is 
the rapid growth with energy of the graviton contribution to the amplitude; 
relative to the pure SM, interference terms go as $\sim s^2/M_s^4$ whereas 
the pure gravity terms behave as $\sim s^4/M_s^8$. For $s$ significantly 
larger than $M_s^2$ tree level unitarity is violated but, long before that, 
we would expect other operators, higher order quantum gravity and stringy 
effects to become important providing some sort of a natural cut off. As one 
can imagine, there are a huge number of processes that one can now examine 
for KK exchange sensitivity and we can only review some of them here. We refer 
the interested reader to the original literature{\cite {pheno2}}.

\section{Lepton Colliders}

Due to the spin-2 nature of the gravitons in the 
tower, angular distributions (and polarization asymmetries) become 
particularly sensitive probes of this 
scenario. For example, the differential cross section for the process 
$e^+e^- \to f\bar f$ now contains both cubic as well as quartic terms in 
$\cos \theta$ and is shown in Fig.2 for $f=b$ at LEP II energies. In all such 
processes the interference between the SM and graviton KK tower exchanges 
is found to vanish when all angles are integrated over thus emphasizing the 
importance of examining differential distributions when trying to constrain 
$M_s$.

\nopagebreak[4]
\begin{figure}[htbp]
\centerline{
\psfig{figure=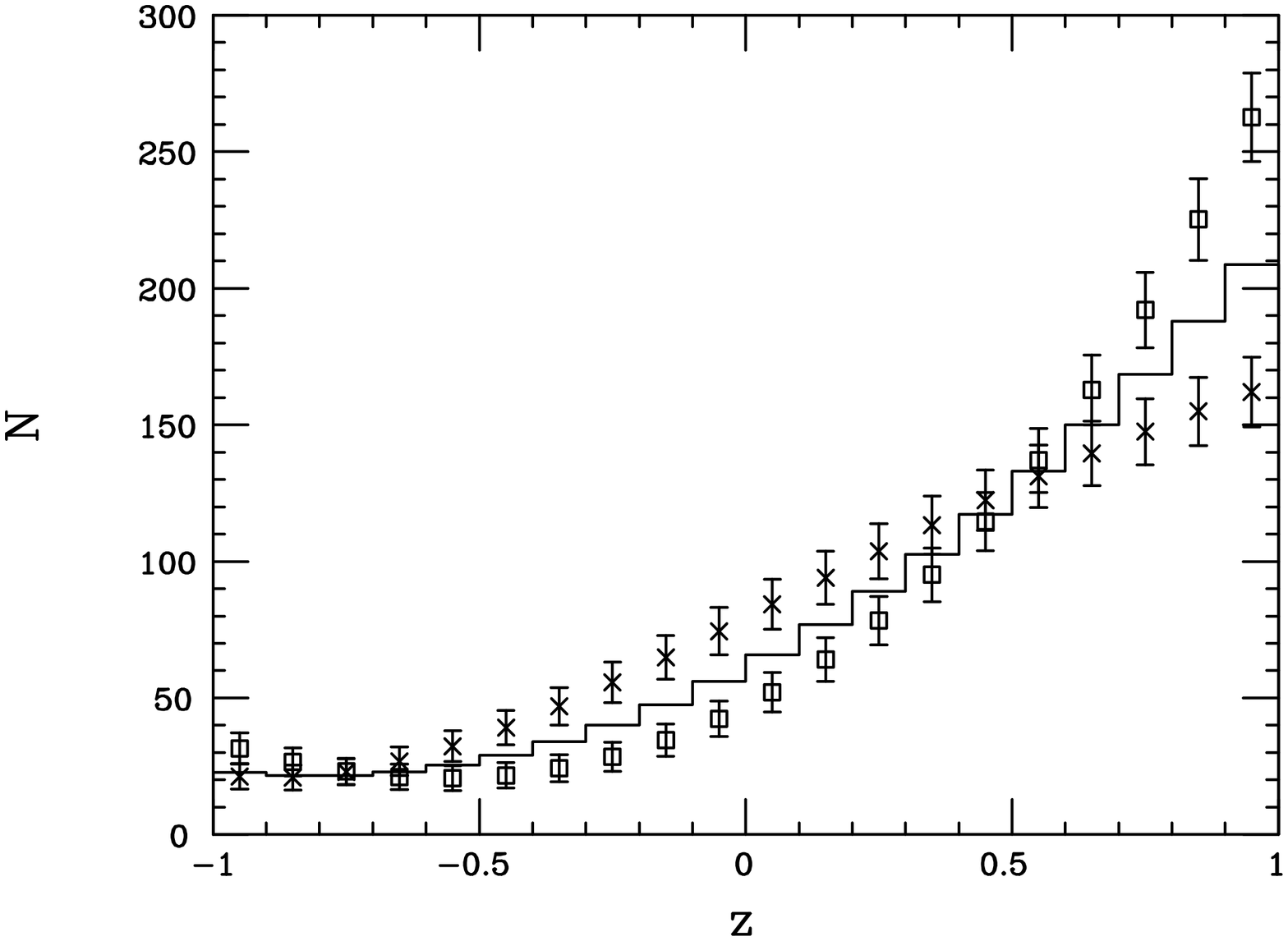,height=6.9cm,width=9cm,angle=-90}}
\vspace*{-.75cm}
\centerline{
\psfig{figure=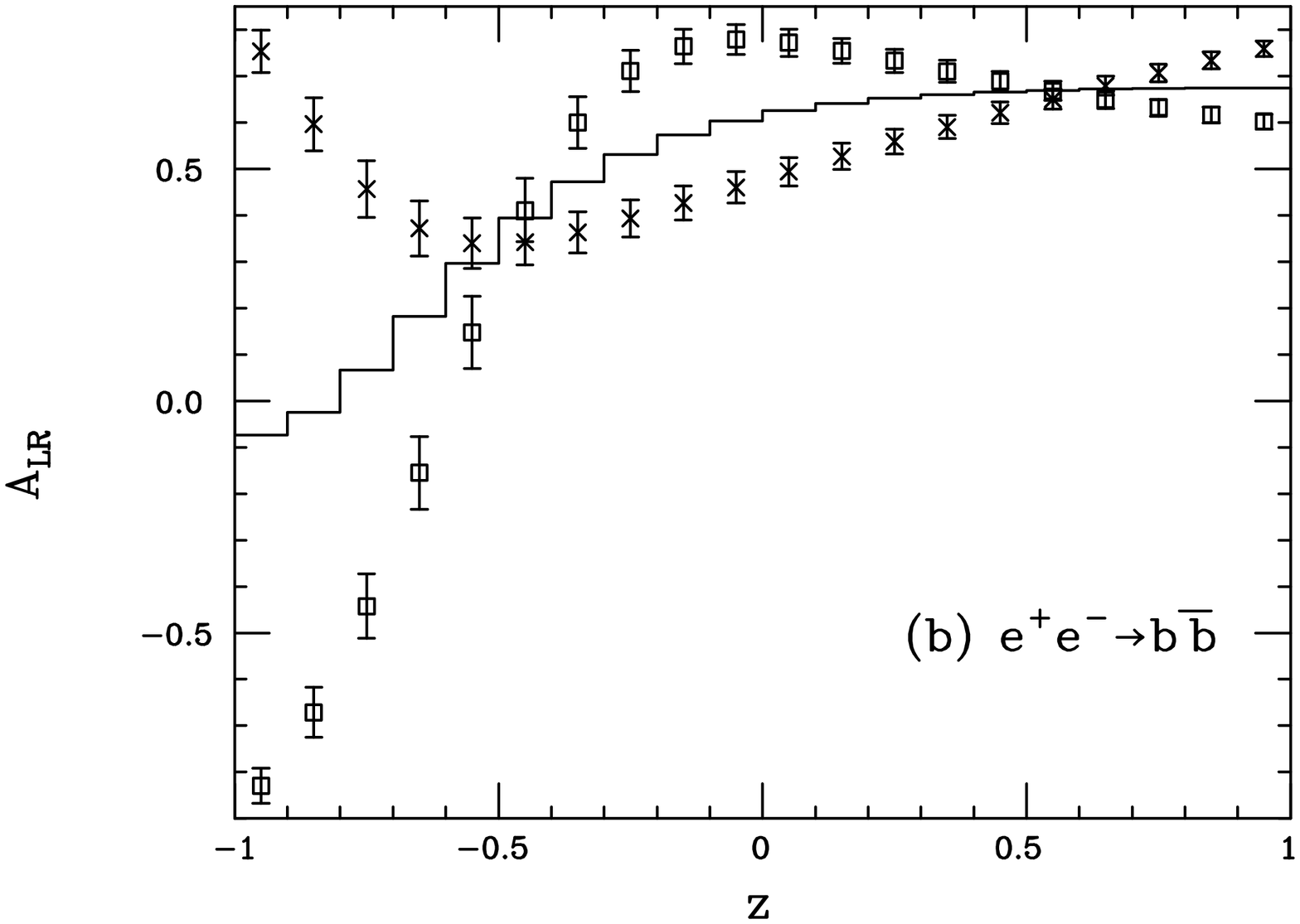,height=6.9cm,width=9cm,angle=-90}}
\vspace*{-0.6cm}
\caption{Distortions(top) in the bin integrated SM cross section(histogram) for 
$e^+e^-\to b\bar b$ at $\sqrt s=200$ GeV with a luminosity of 1 $fb^{-1}$ 
assuming $M_s=0.6$ TeV. The two sets of `data' correspond to $\lambda=\pm 1$.
The corresponding distortions in the $b$ Left-Right Asymmetry at a 500 GeV 
linear collider are shown in the bottom panel assuming a luminosity of 75 
$fb^{-1}$.}
\end{figure}
\nopagebreak[4]

Hewett{\cite {pheno1}} performed a combined fit of the angular distributions 
of all kinematically accessible $f\bar f$ final states, as well as the 
polarization of the $\tau$, to obtain a potential search reach for LEP through 
these channels of $M_s \simeq 1$ TeV for either sign of $\lambda$. She also 
demonstrated that $e^+e^-$ colliders could distinguish the particular 
deviations 
induced by these spin-2 exchanges from those due to lower spins, such as a 
$Z'$ or $\tilde \nu$ in R-parity violating models{\cite {rp}, or from 
exchanges in other channels, such as leptoquarks{\cite {lepto}}, almost up to 
the search reach. Such an analysis can be extended to the case of a 500 GeV or 
1 TeV 
linear collider with the results as shown in Fig.3. Here not only are the 
differential cross sections employed but also the angular dependent 
polarization asymmetries, shown in Fig.2 for the case of the $b$ quark, 
since the initial electron beam can now be polarized. 
We see that the reach from such a combined channel analysis of this type is of 
order $M_s \sim 6-7 \sqrt s$. 

\nopagebreak[4]
\begin{figure}[htbp]
\centerline{
\psfig{figure=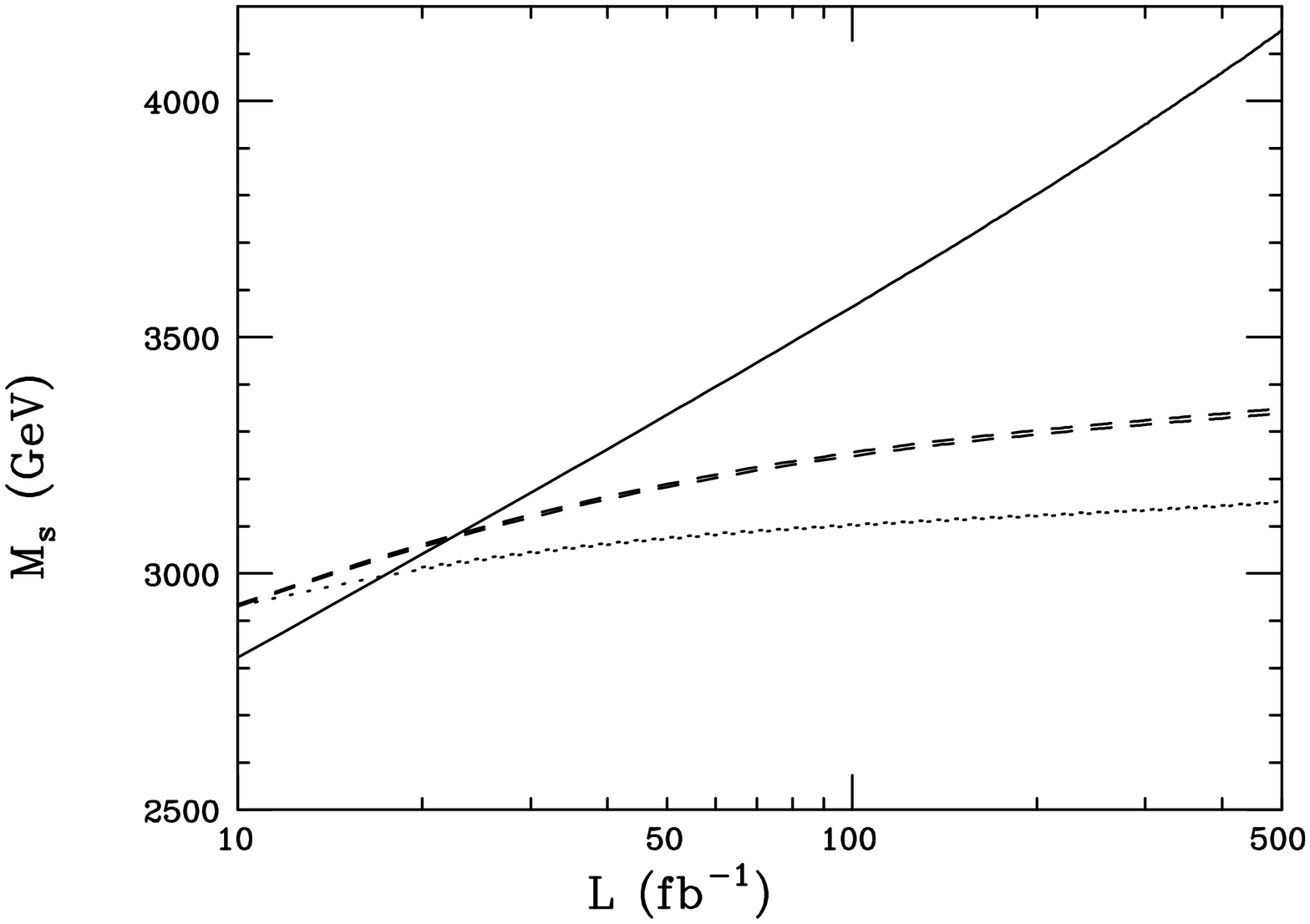,height=6.9cm,width=9cm,angle=-90}}
\vspace*{-.75cm}
\centerline{
\psfig{figure=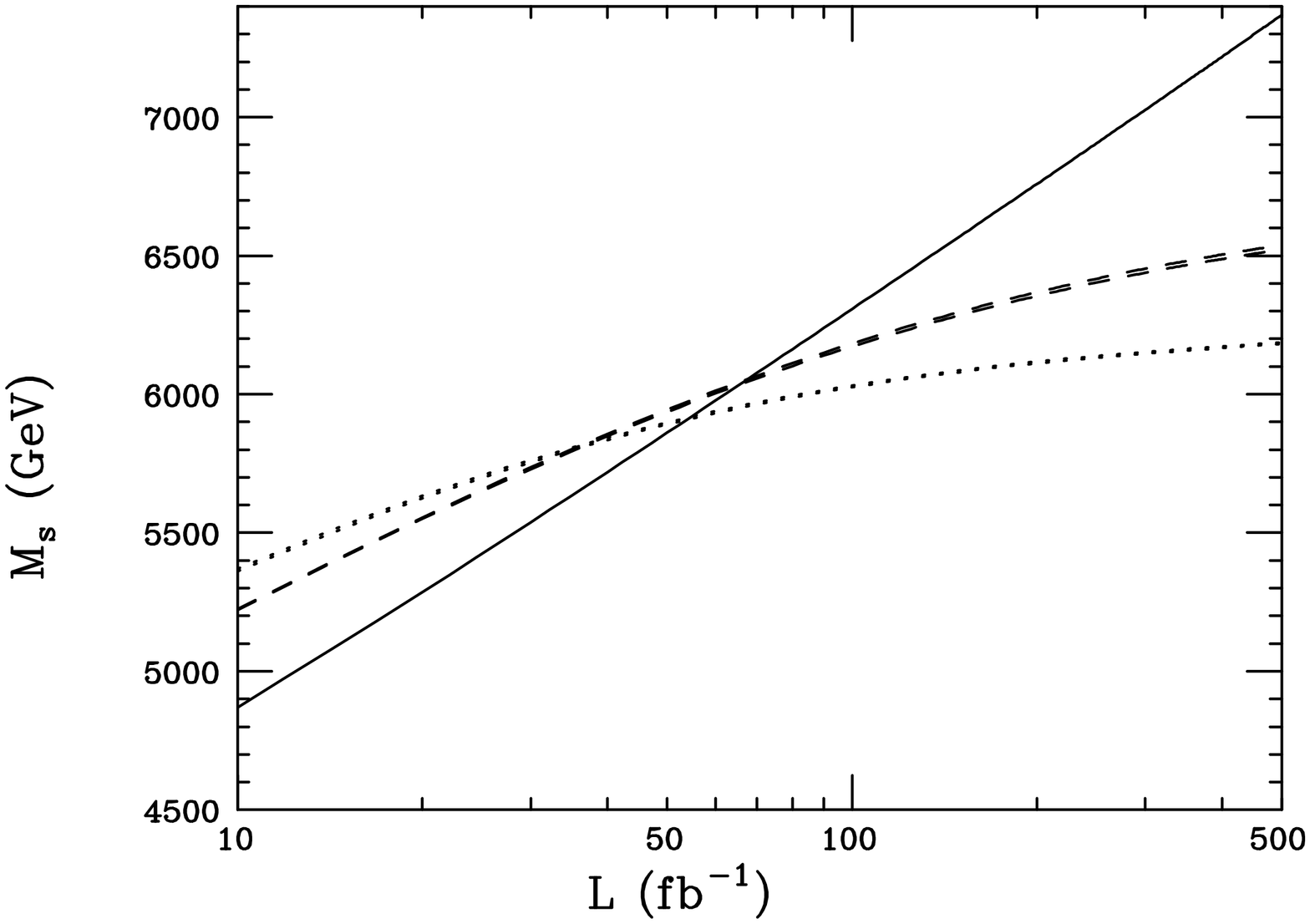,height=6.9cm,width=9cm,angle=-90}}
\vspace*{-0.6cm}
\caption{Search reaches for $M_s$ at a 500 GeV(top) and 1000 GeV(bottom) 
$e^+e^-/e^-e^-$ collider as 
a function of the integrated luminosity for Bhabha(dashed) and Moller(dotted) 
scattering for either sign of the parameter $\lambda$ in comparison to the 
`usual' search employing the combined channel 
$e^+e^-\to f\bar f$(solid) fit as described in the text.}
\end{figure}
\nopagebreak[4]

$e^+e^-$ annihilation into gauge boson pairs also can provide reasonable 
sensitivity to $M_s$ as shown in Figs.4 and 5 for $W^+W^-$ and $ZZ$ 
production at LEP II, respectively. We note that unlike what happens in 
many other new physics 
scenarios, the $W^+W^-$ total cross section can actually be {\it decreased} 
through the exchange of KK towers of gravitons. While the KK towers do not 
lead to any appreciable modification of the $ZZ$ angular distribution, they 
modify the $W^+W^-$ distribution in the backwards direction due to the 
dominance of the SM $t-$channel $\nu$ exchange graph. The KK tower exchange is 
found to lead to essentially insignificant modifications in the helicity 
fractions of the final state gauge bosons at LEP II energies although the 
effects are somewhat larger at higher energy colliders{\cite {pheno2}}. The 
differential cross section 
modifications in the case of the $\gamma\gamma$ final state are similar to 
those of $ZZ$. 

\nopagebreak[4]
\begin{figure}[htbp]
\centerline{
\psfig{figure=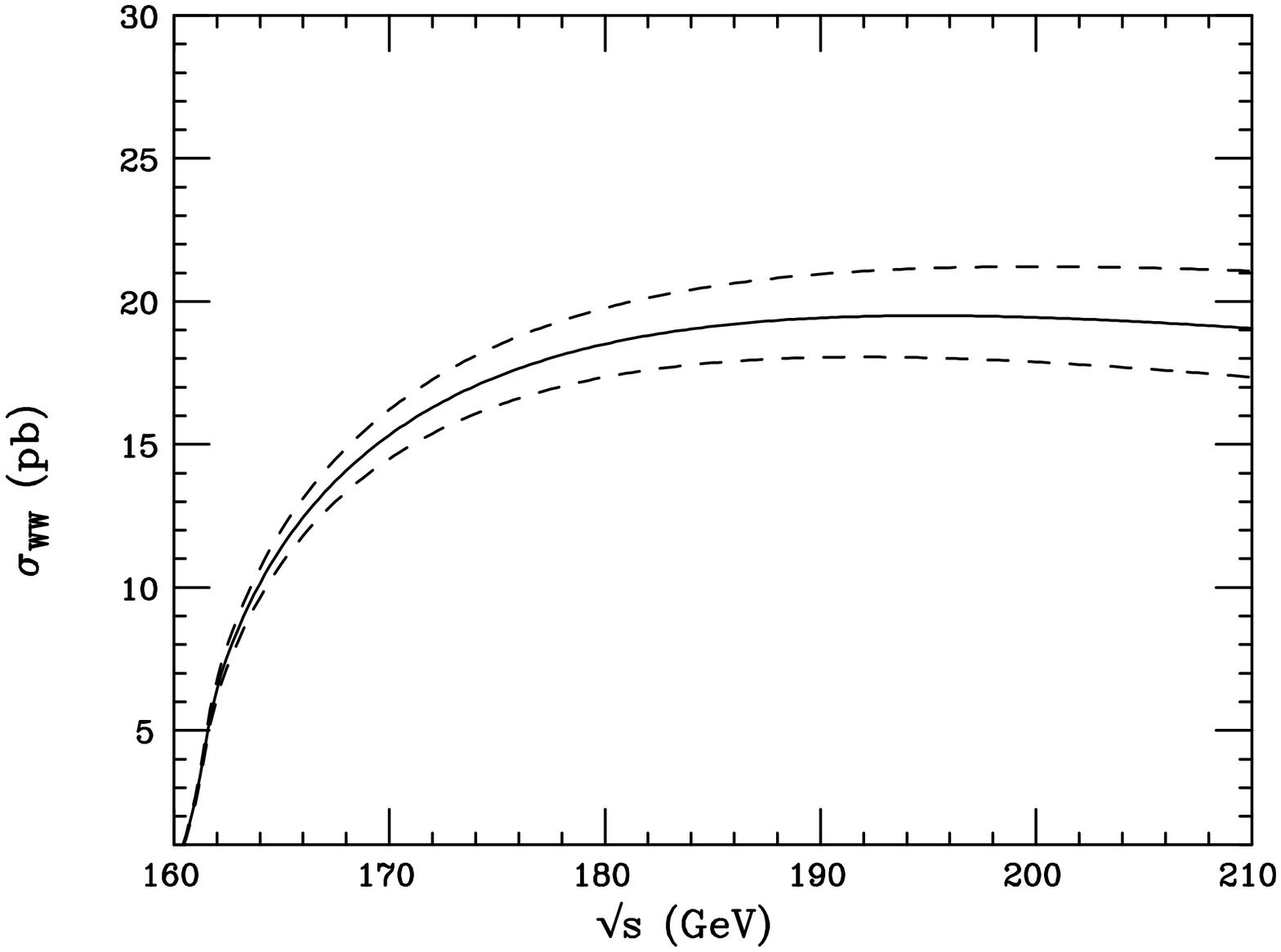,height=6.9cm,width=9cm,angle=-90}}
\vspace*{-.75cm}
\centerline{
\psfig{figure=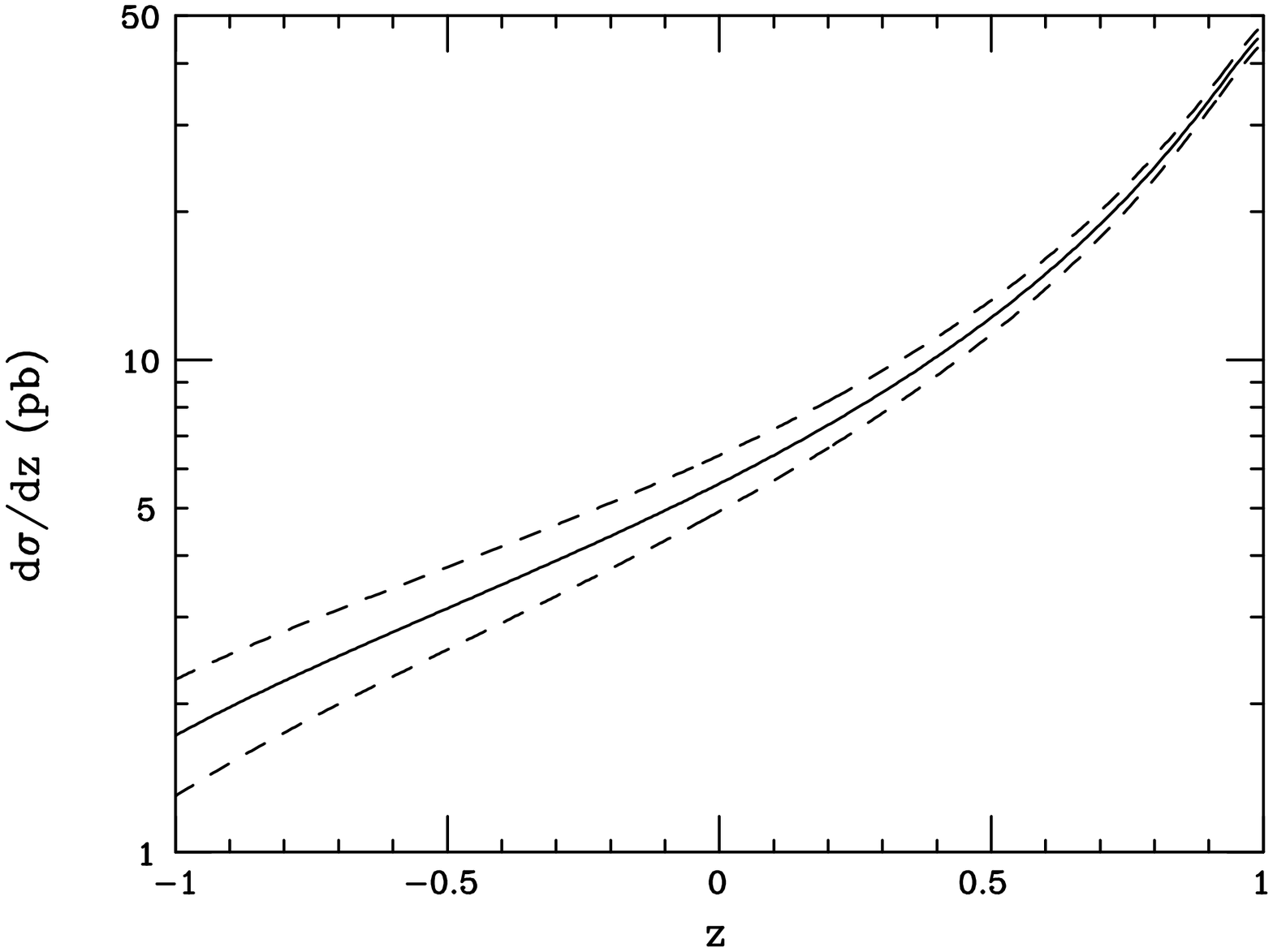,height=6.9cm,width=9cm,angle=-90}}
\vspace*{-0.6cm}
\caption{Total cross section and angular distribution for $W^+W^-$ production 
at LEP II energies for the case of the SM (solid) or the ADD model with 
$M_s=0.6$ TeV and $\lambda=\pm 1$(dashed). The angular distribution in the 
lower panel is for $\sqrt s=200$ GeV. Note $z=\cos \theta$.}
\end{figure}
\nopagebreak[4]

As mentioned earlier, graviton exchange can lead to new processes which are 
forbidden at the tree level in the SM. An excellent example of this is 
provided by the process $e^+e^- \to 2h$ where $h$ is a Higgs boson; in the SM 
or MSSM this process can only occur at the one-loop level. The angular 
distribution induced from the tree level graviton exchange and that arising 
from loops are quite distinct since the graviton exchange graph leads to a 
distribution quartic in $\cos \theta$. The cross section 
is reasonably large with measurable rates possible at future linear colliders 
for $M_s \sim 1$ TeV but are found to decrease rapidly as $M_s$ is 
increased, \ie,  $\sigma \sim s^3/M_s^8$ as shown in Fig.6. 

\nopagebreak[4]
\begin{figure}[htbp]
\centerline{
\psfig{figure=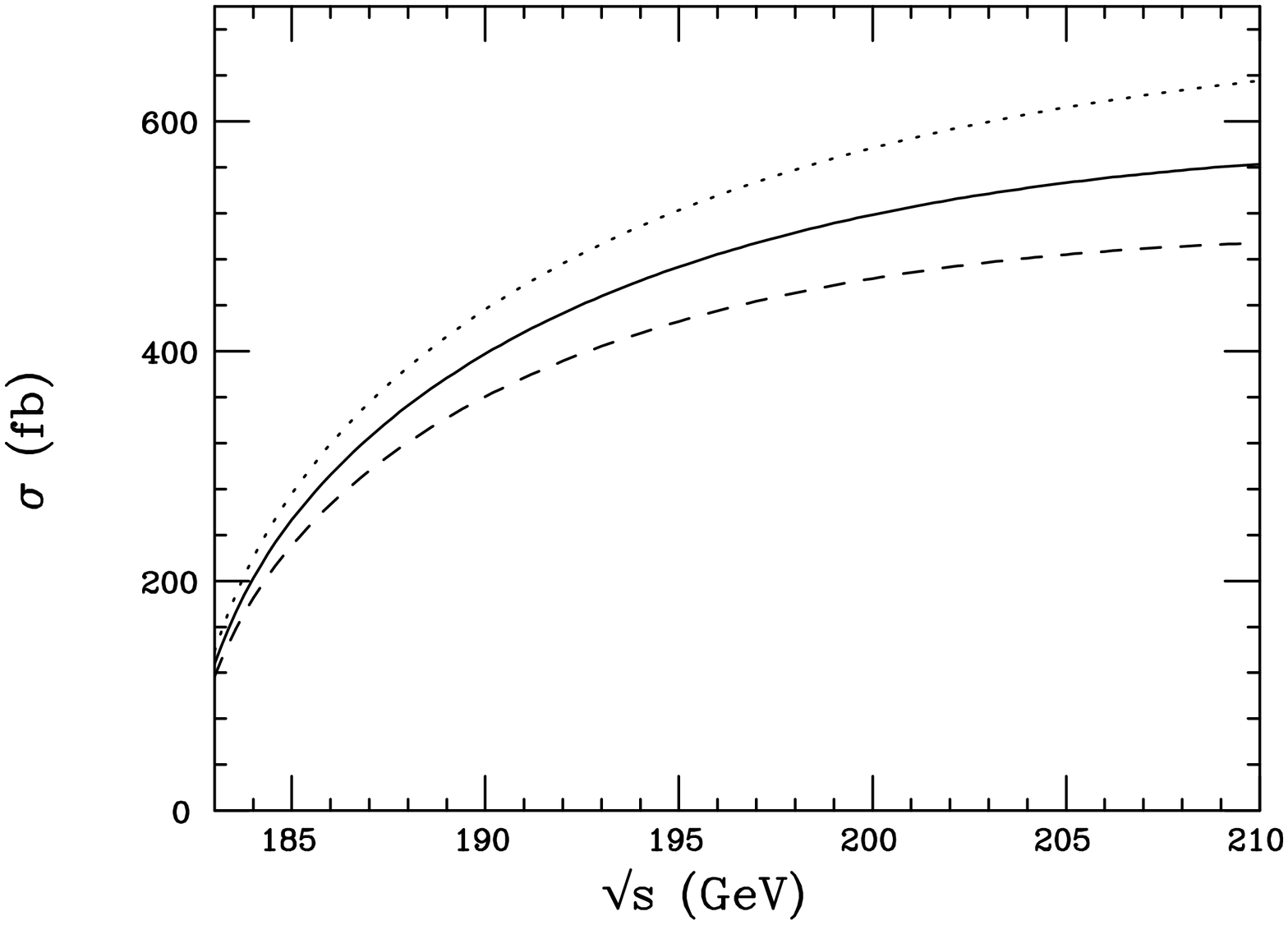,height=6.9cm,width=9cm,angle=-90}}
\vspace*{-.75cm}
\centerline{
\psfig{figure=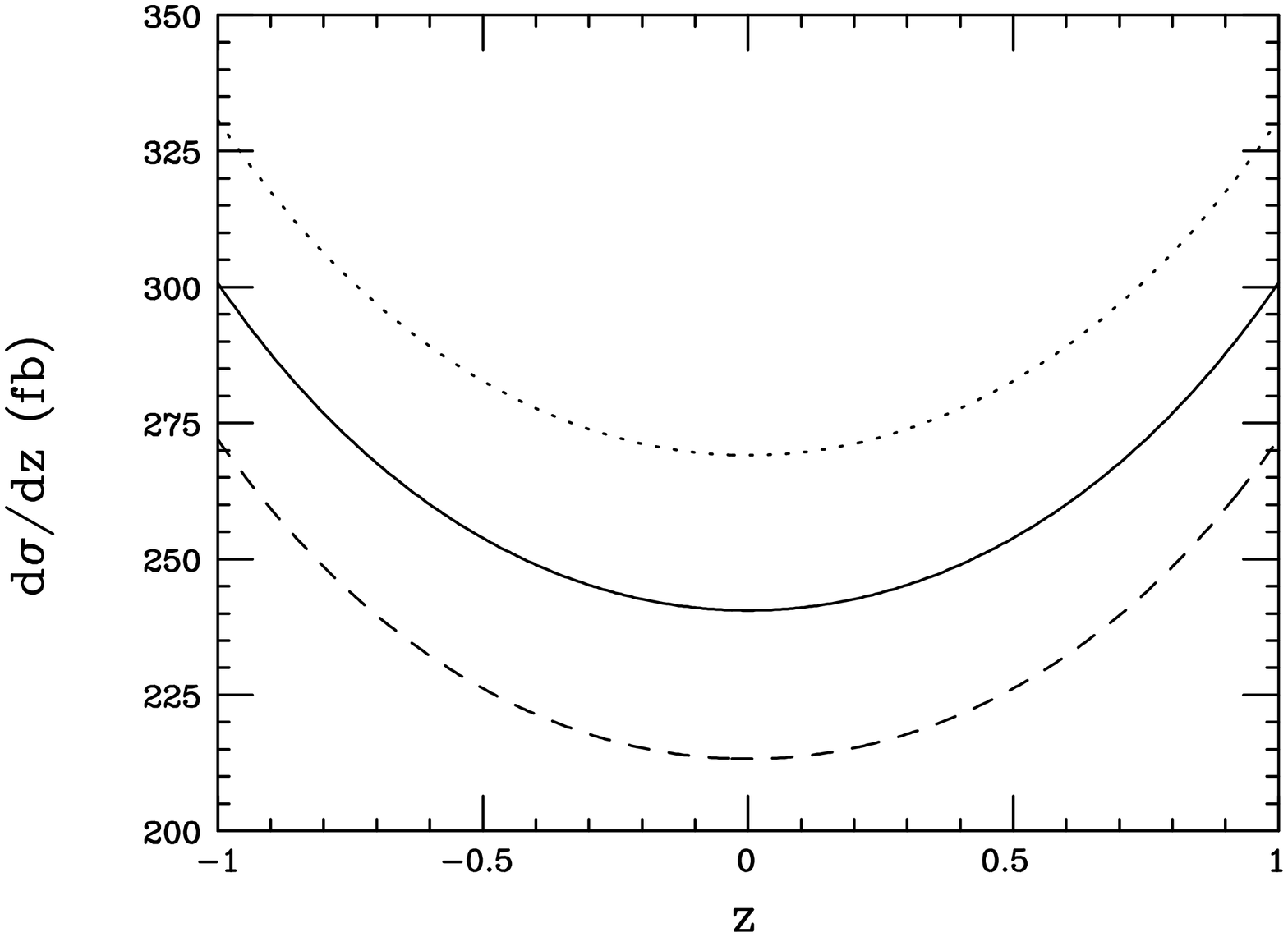,height=6.9cm,width=9cm,angle=-90}}
\vspace*{-0.6cm}
\caption{Same as in the previous figure but now for the $ZZ$ final state.}
\end{figure}
\nopagebreak[4]

\nopagebreak[4]
\vspace*{-0.5cm}
\nn
\begin{figure}[htbp]
\centerline{
\psfig{figure=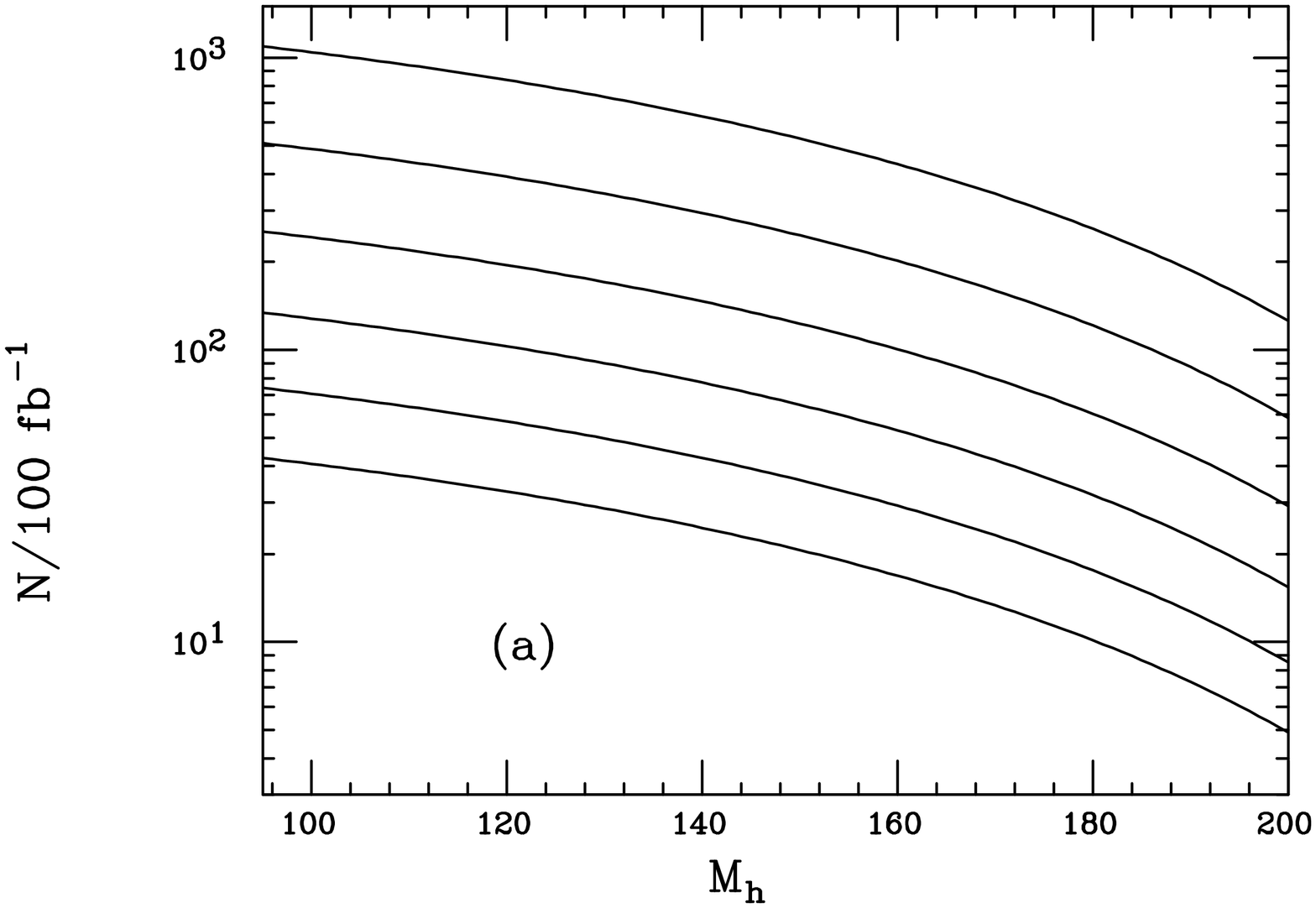,height=8.0cm,width=9cm,angle=-90}}
\vspace*{-0.6cm}
\caption{Tree level production rate for Higgs boson pairs due to graviton tower 
exchange at a 500 GeV $e^+e^-$ 
collider as a function of the Higgs mass scaled to an integrated luminosity 
of 100 $fb^{-1}$. From top to bottom the curves correspond to the 
choice $M_s=1$ TeV increasing in steps of 0.1 TeV.}
\end{figure}
\vspace*{0.1mm}

\section{Hadron Colliders}

Hadron colliders such as the Tevatron and LHC offer many distinctive probes 
for the exchange of KK towers of 
gravitons. Here we will limit our discussion to the Drell-Yan process and top 
quark pair production.

\nopagebreak[4]
\vspace*{-0.5cm}
\nn
\begin{figure}[htbp]
\centerline{
\psfig{figure=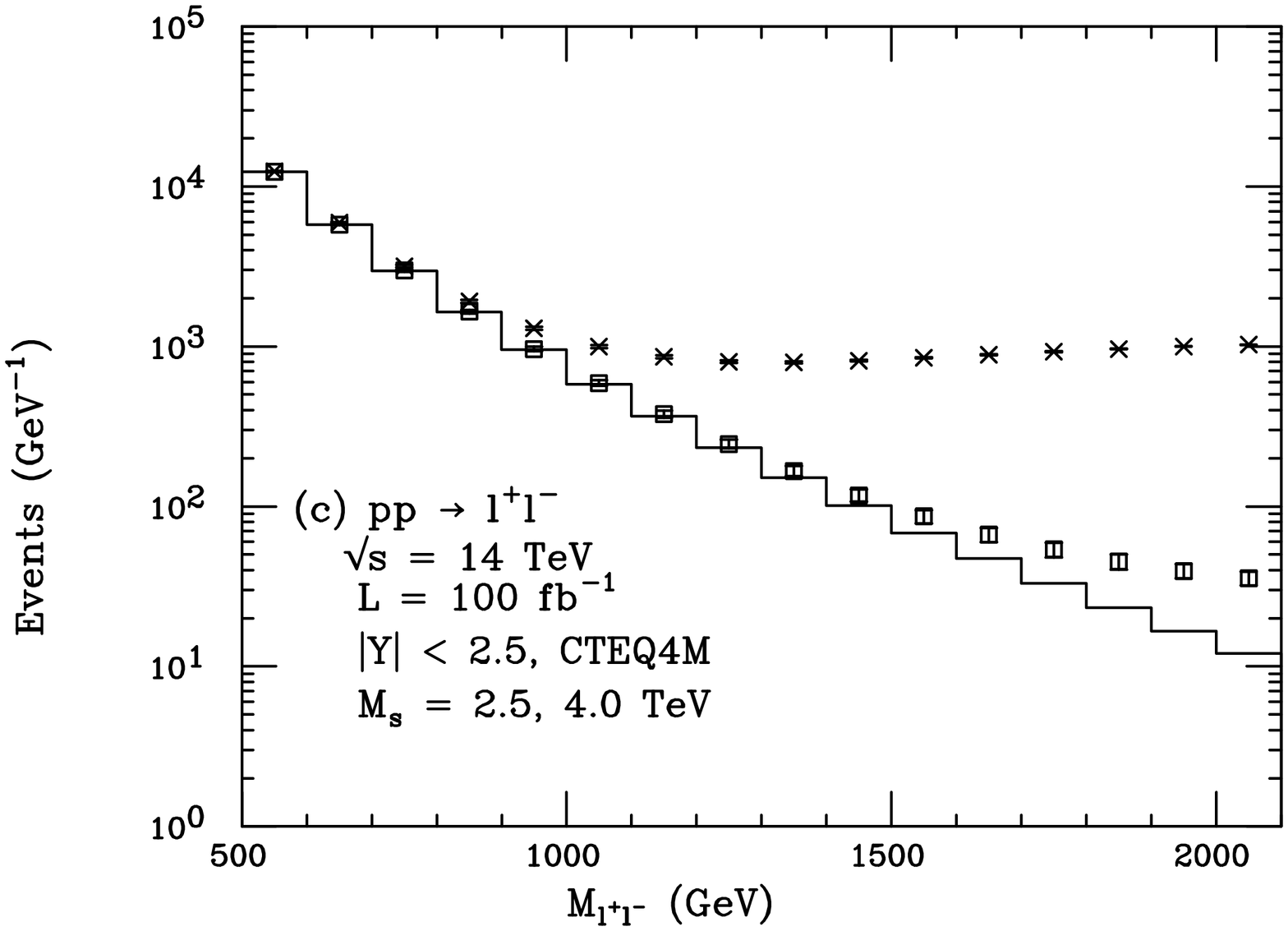,height=6.9cm,width=9cm,angle=-90}}
\vspace*{-10mm}
\centerline{
\psfig{figure=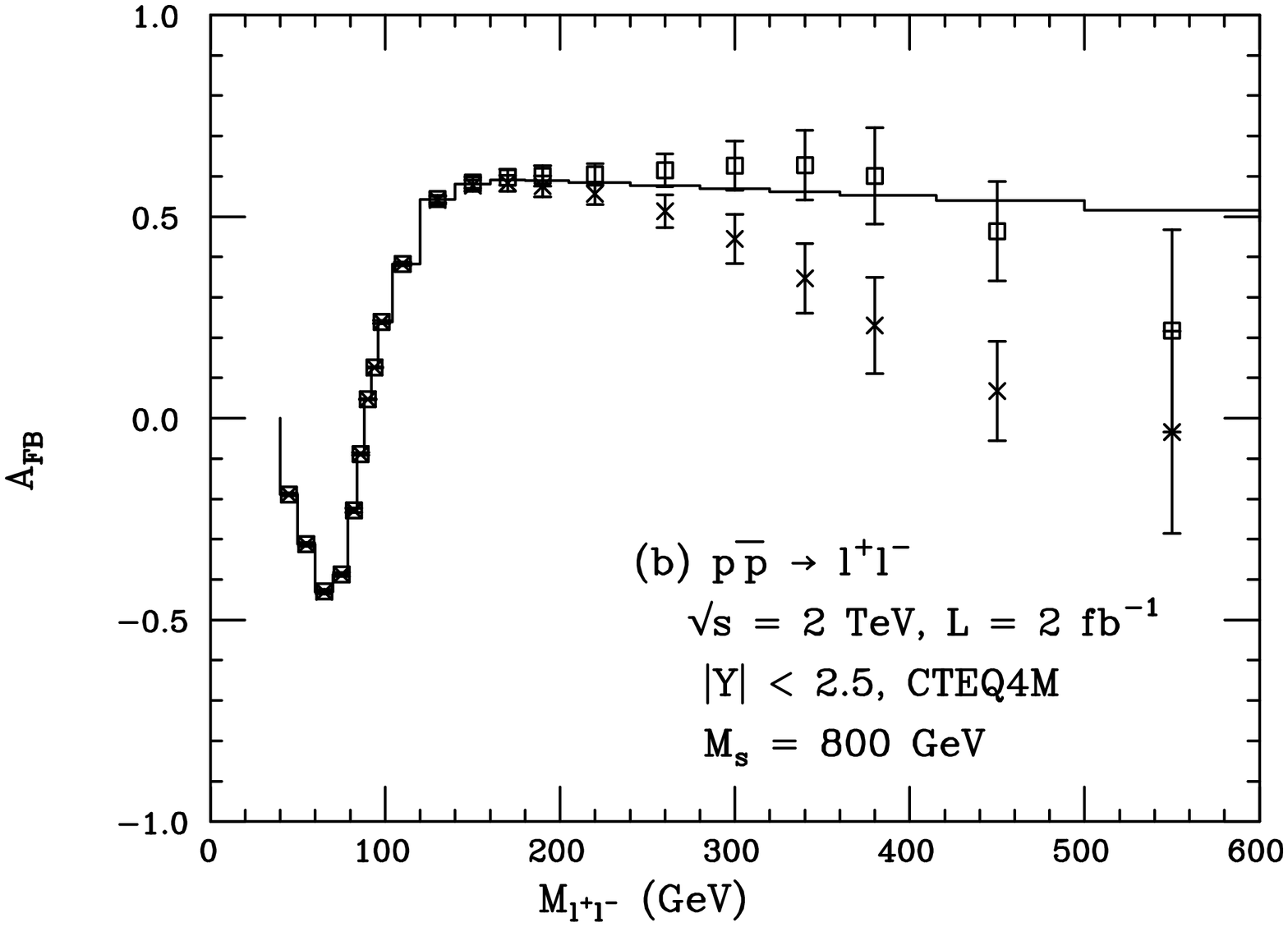,height=6.9cm,width=9cm,angle=-90}}
\vspace*{-0.9cm}
\caption{Drell-Yan cross section at the LHC in the SM(histogram) as well as 
for $M_s$=2.5 or 4 TeV. $A_{FB}$ at the Tevatron with $M_s$=0.8 TeV for either 
sign of $\lambda$.}
\end{figure}
\nopagebreak[4]

An interesting feature of the ADD scenario 
in the case of Drell-Yan channel is the contribution 
of $gg$ initial state to the charged dilepton pair production, \ie, 
$gg,q\bar q \to G_n \to \ell^+\ell^-$. This additional amplitude helps to 
increase the $M_s$ reach for this process, particularly at the LHC where the 
$gg$ luminosities are so large. As in the case of $e^+e^-\to f\bar f$, the KK 
exchange modifies both the cross section as well as the angular distribution 
as probed by the Forward-Backward asymmetry, $A_{FB}$. In the case of the 
total cross section as the invariant pair mass of the leptons approaches $M_s$ 
there is a substantial rise in the cross section, for either sign of 
$\lambda$, appearing 
in the form of a shoulder as is typical of contact interaction 
signatures. Similarly as the scale $M_s$ is approached from below the 
magnitude of $A_{FB}$ is seen to decrease towards zero. This is easily 
understood since the graviton KK tower couplings are $C$ and $P$ conserving 
thus leading to a angular distribution for leptons which involve only even 
powers of $\cos \theta$ and a null asymmetry. Fig.7 shows how these deviations 
would appear at hadron colliders. Hewett{\cite {pheno1}} has estimated the 
$M_s$ search reach for the Drell-Yan channel at both the Tevatron and LHC by 
performing a fit to both the bin integrated total cross section and $A_{FB}$ 
and obtained the results shown in Fig.8. In particular we see that the LHC 
reach is $\simeq 5.3$ TeV, somewhat less than that obtainable at a 1 TeV 
linear collider. 

\nopagebreak[4]

\vspace*{-0.5cm}
\nn
\begin{figure}[htbp]
\centerline{
\psfig{figure=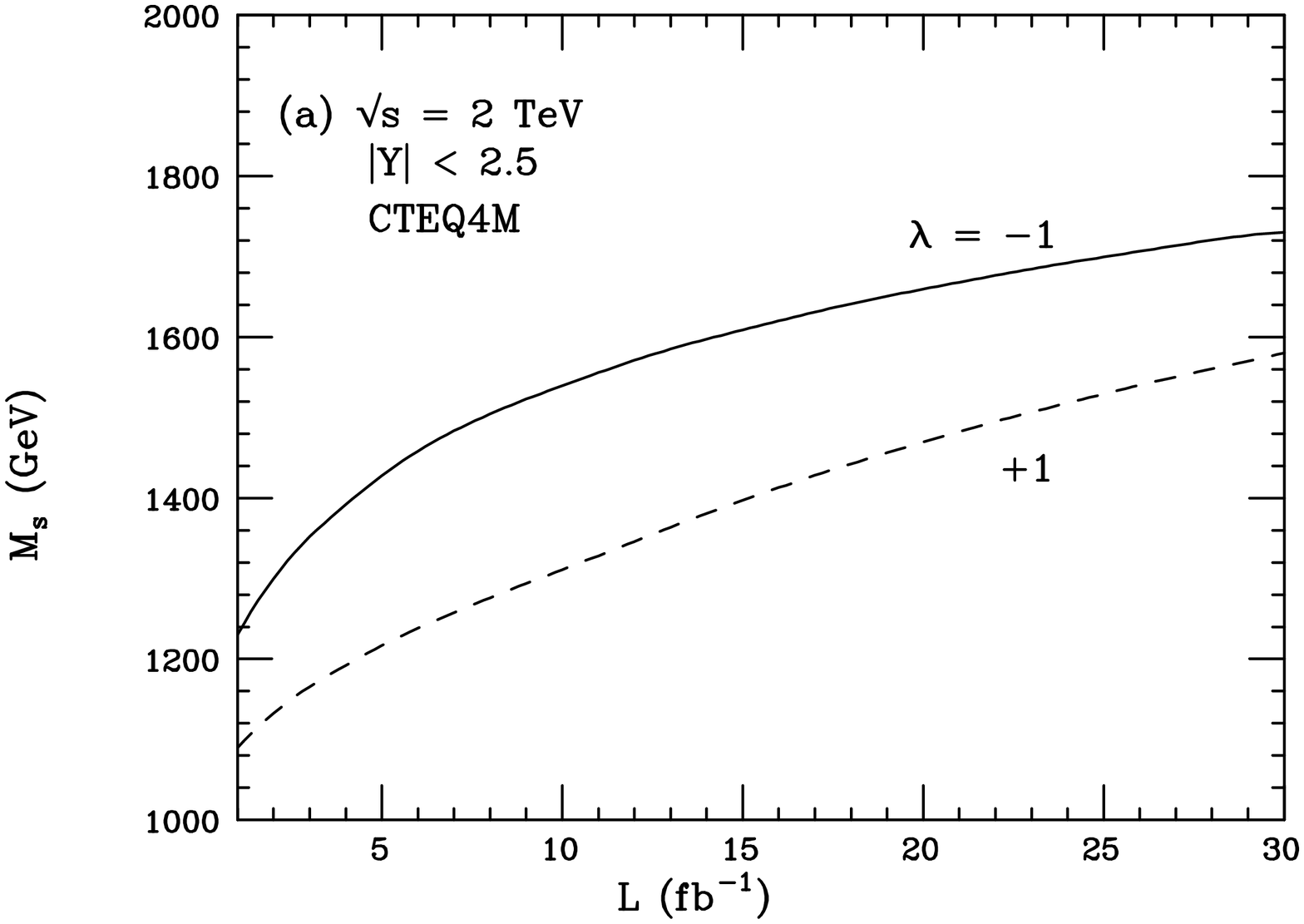,height=6.9cm,width=9cm,angle=-90}}
\vspace*{-10mm}
\centerline{
\psfig{figure=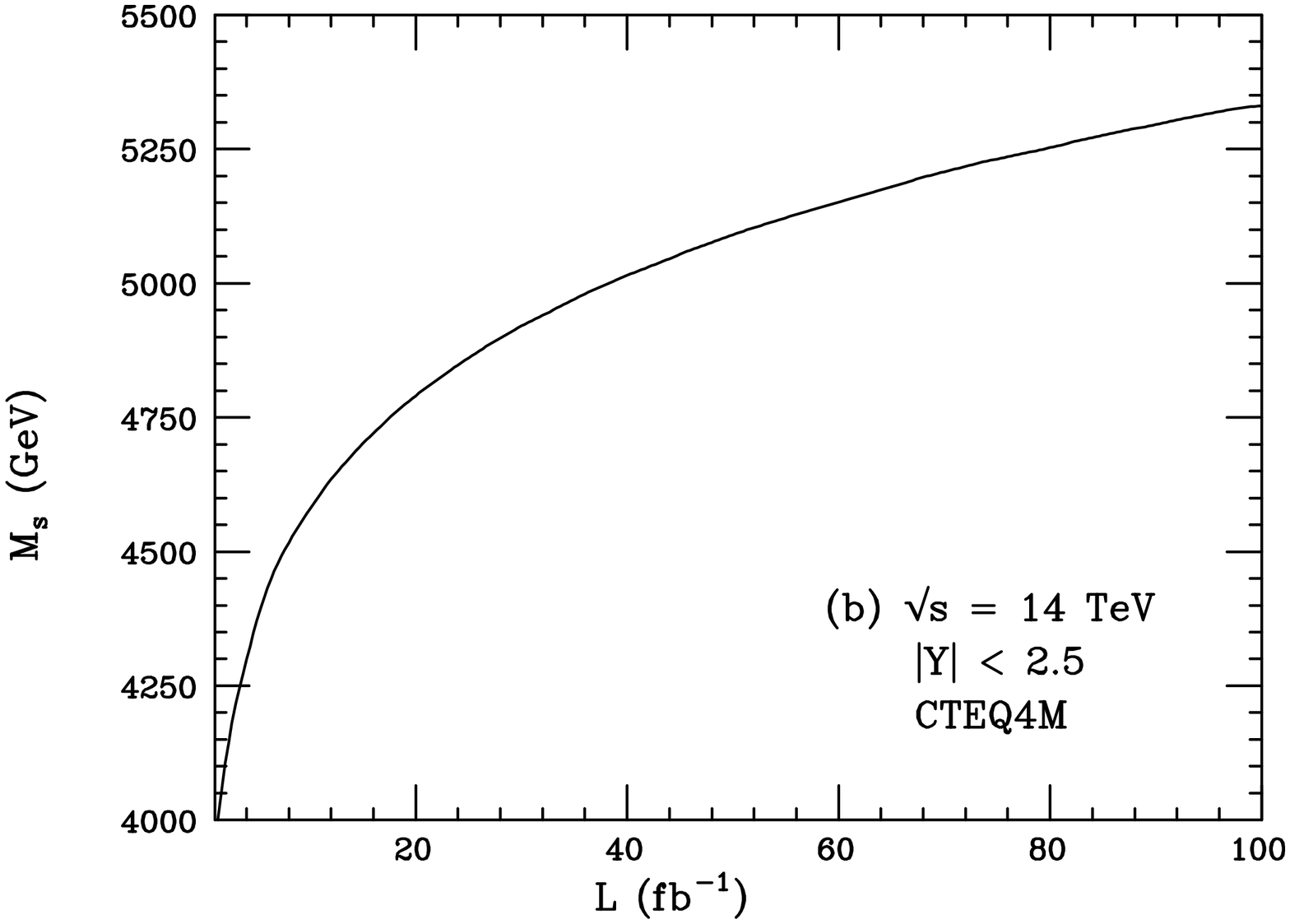,height=6.9cm,width=9cm,angle=-90}}
\vspace*{-0.9cm}
\caption{$M_s$ reaches at the Tevatron and LHC from Drell-Yan.}
\end{figure}
\nopagebreak[4]

In the case of top pair production it has been shown that the deviation in 
the cross section due to graviton tower exchanges, 
$gg,q\bar q \to G_n \to t\bar t$, 
is not the best probe for large values of $M_s$. (The respective 
deviations in the total $t\bar t$ cross sections at the Tevatron and LHC due 
to KK gravitons are shown in Figs.9 and 10.) Thus in order to gain in 
sensitivity we must look at the various distributions. It{\cite {pheno2}} 
has also been shown that neither the rapidity nor the top quark 
angular distributions are 
particularly sensitive to these new contributions. However, the top quark 
$p_t$ and top pair invariant mass distributions do show substantial sensitivity 
to finite $M_s$ as is also shown in Figs. 9 and 10. Performing a simultaneous 
fit to both the $p_t$ and $M_{t\bar t}$ distributions leads to a reach of 
1.2(6) TeV at the Run II Tevatron with an integrated luminosity of 2 $fb^{-1}$ 
(LHC with an integrated luminosity of 
100 $fb^{-1}$). These values are quite comparable to but somewhat better than 
those obtained through the analysis of the Drell-Yan channel.

\nopagebreak[4]
\vspace*{-0.5cm}
\nn
\begin{figure}[htbp]
\centerline{
\psfig{figure=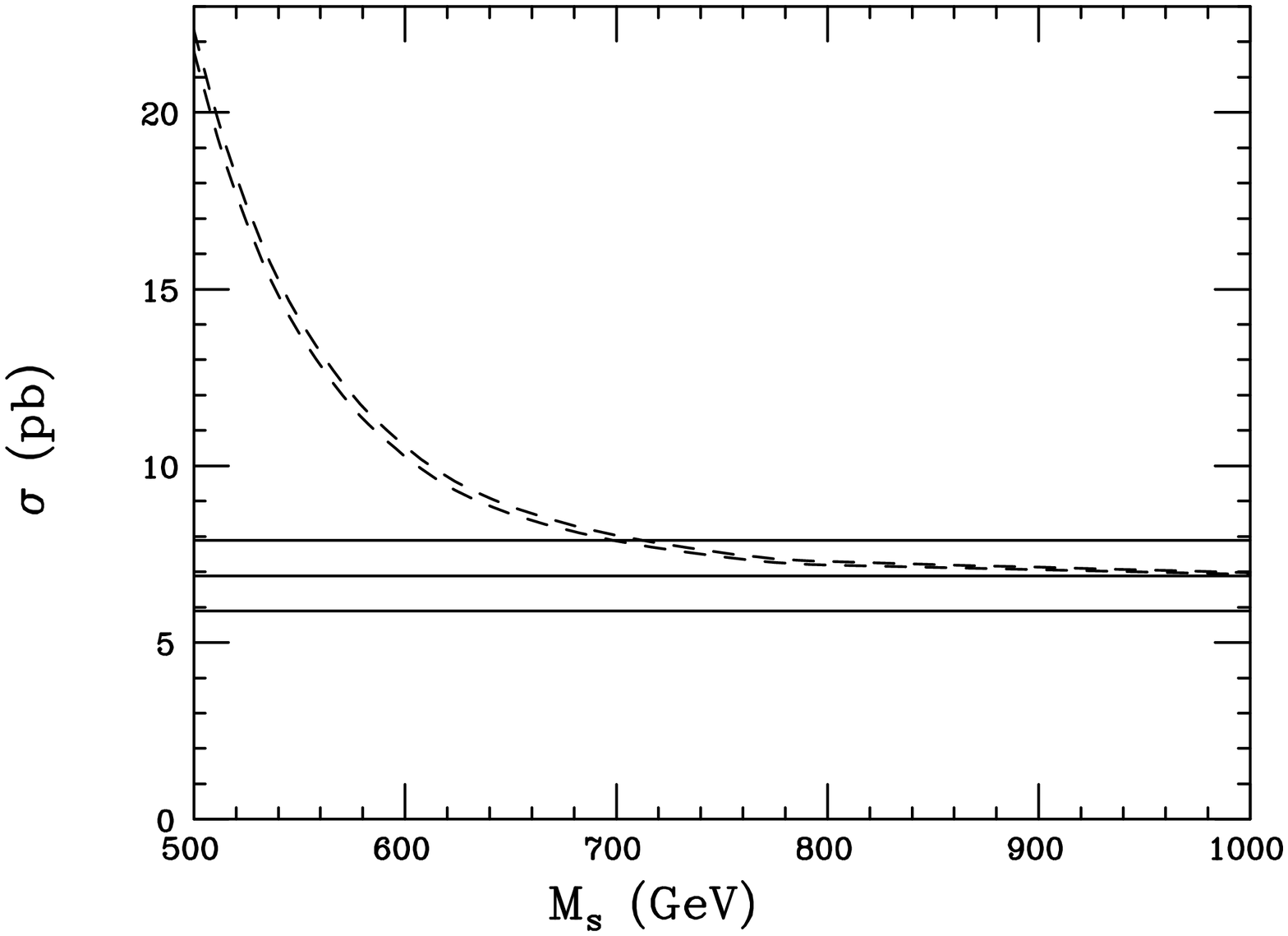,height=6.9cm,width=9cm,angle=-90}}
\vspace*{-10mm}
\centerline{
\psfig{figure=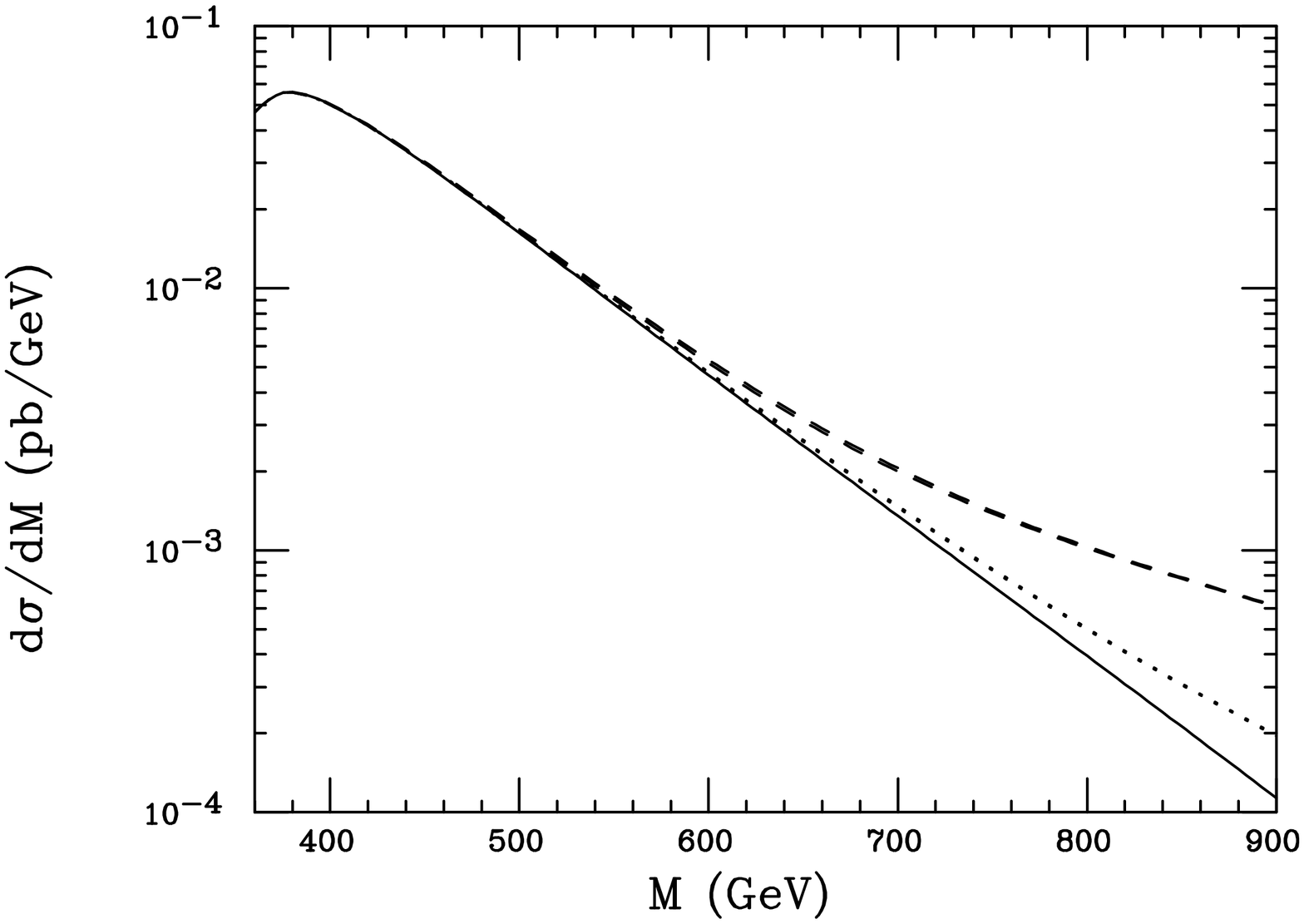,height=6.9cm,width=9cm,angle=-90}}
\vspace*{-0.9cm}
\caption{Top pair production cross section at Run II of the Tevatron as a 
function of $M_s$. The solid band represents an approximate 
$15\%$ error on the cross section determination. Although the results for 
both $\lambda=\pm 1$ are shown they are visually difficult to separate. Also 
shown is the top pair invariant mass distribution at Run II of the Tevatron 
assuming the SM(solid) or $M_s=800(1000)$ GeV as the dashed(dotted) curve.}
\end{figure}
\nopagebreak[4]

\section{HERA}

$ep$ collisions at HERA allow us to probe $t$-channel graviton tower exchange 
via the processes $eq,g \to eq,g$; note that the gluon contribution in now 
present as it was for the Drell-Yan process. By fitting the cross sections 
at low-$Q^2$ we can extrapolate the cross sections 
to larger $Q^2$ to obtain an estimate of 
the $M_s$ reach. This result is found to be fairly insensitive to the choice 
of beam polarization or whether $e^+$ or $e^-$ beams are employed. The reach 
as a function of the integrated luminosity is shown in Fig.11{\cite {pheno2}}.

\nopagebreak[4]
\vspace*{-0.5cm}
\nn
\begin{figure}[htbp]
\centerline{
\psfig{figure=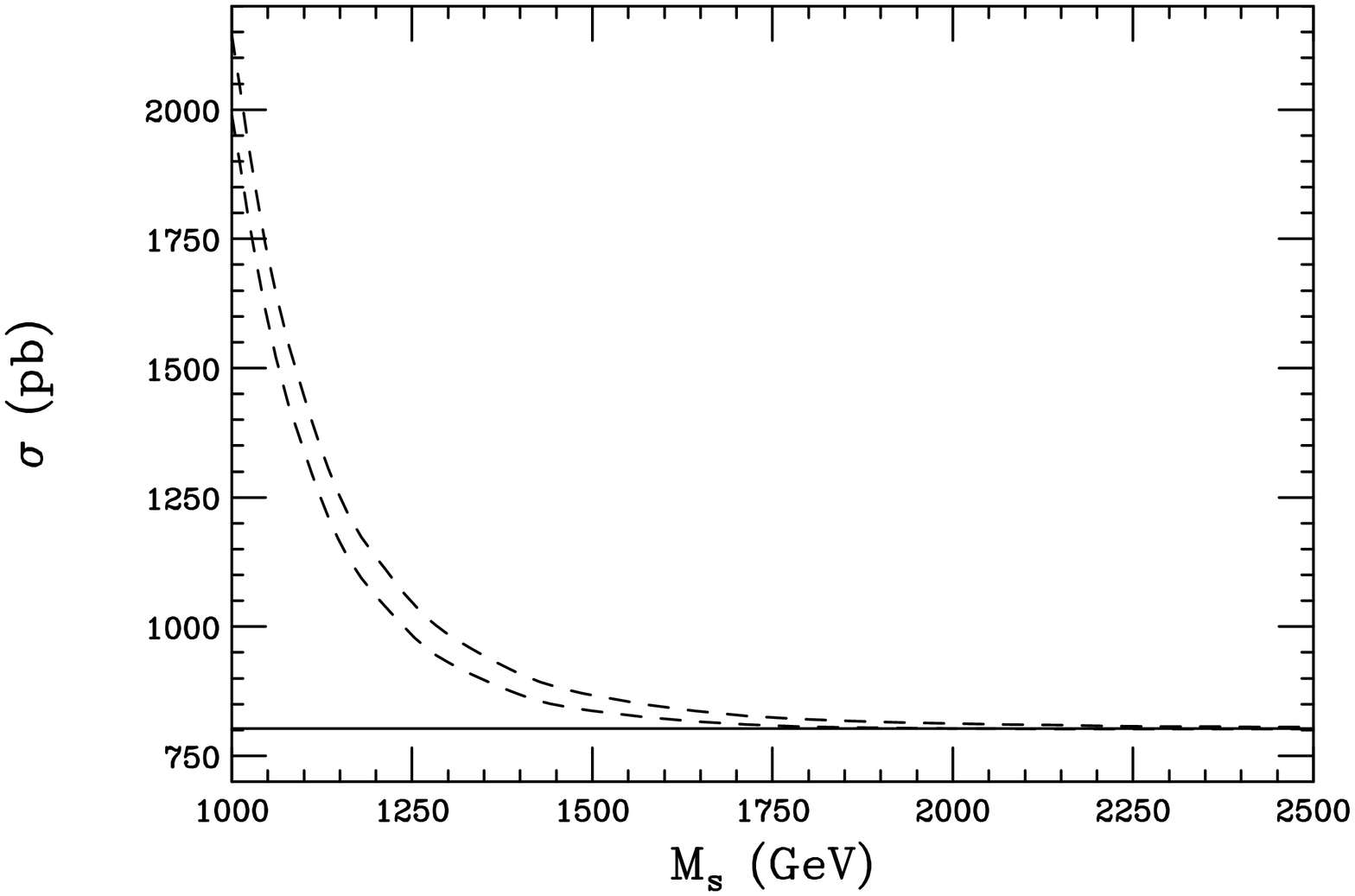,height=5.9cm,width=8cm,angle=90}}
\vspace*{0.9cm}
\centerline{
\psfig{figure=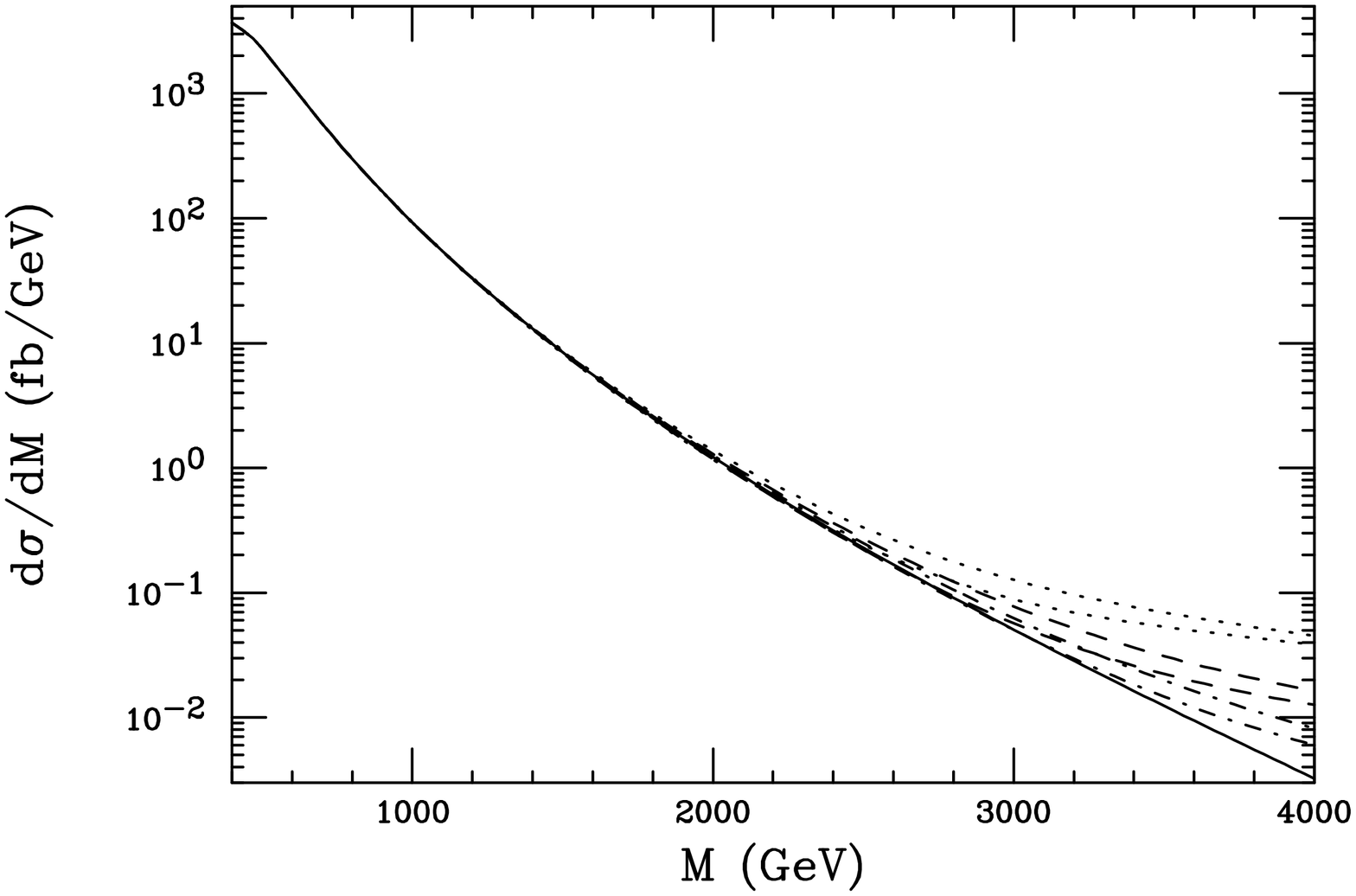,height=5.9cm,width=8cm,angle=90}}
\vspace*{0.8cm}
\caption{Top pair production cross section and pair invariant mass 
distribution at LHC. 
In the top quark pair invariant mass plot the values of $M_s$ were chosen to 
be 3, 3.5 and 4 TeV with $\lambda=\pm 1$.}
\end{figure}
\nopagebreak[4]

\section{$\gamma \gamma$ and $\gamma e$ Colliders}

Polarized $\gamma\gamma$ and $\gamma e$ collisions may be possible at future 
$e^+e^-$ 
colliders through the use of Compton backscattering of polarized low energy 
laser beams off of polarized high energy electrons. The resulting 
backscattered photon distribution, $f_\gamma(x=E_\gamma/E_e)$, is far from 
monoenergetic and is cut off above $x_{max}\simeq 0.83$ implying that 
the colliding photons are significantly softer than the parent lepton beam 
energy. This cutoff at 
large $x$ implies that the $\gamma \gamma(e)$ center of mass energy 
never exceeds $\simeq 0.83(0.91)$ of the parent collider. In 
addition, both the shape of the function $f_\gamma$ and the average helicity 
of the produced $\gamma$'s are quite sensitive to the 
polarization state of both the initial laser ($P_l$) and electron ($P_e$) 
whose values fix the specific distribution.

\nopagebreak[4]
\vspace*{-0.5cm}
\nn
\begin{figure}[htbp]
\centerline{
\psfig{figure=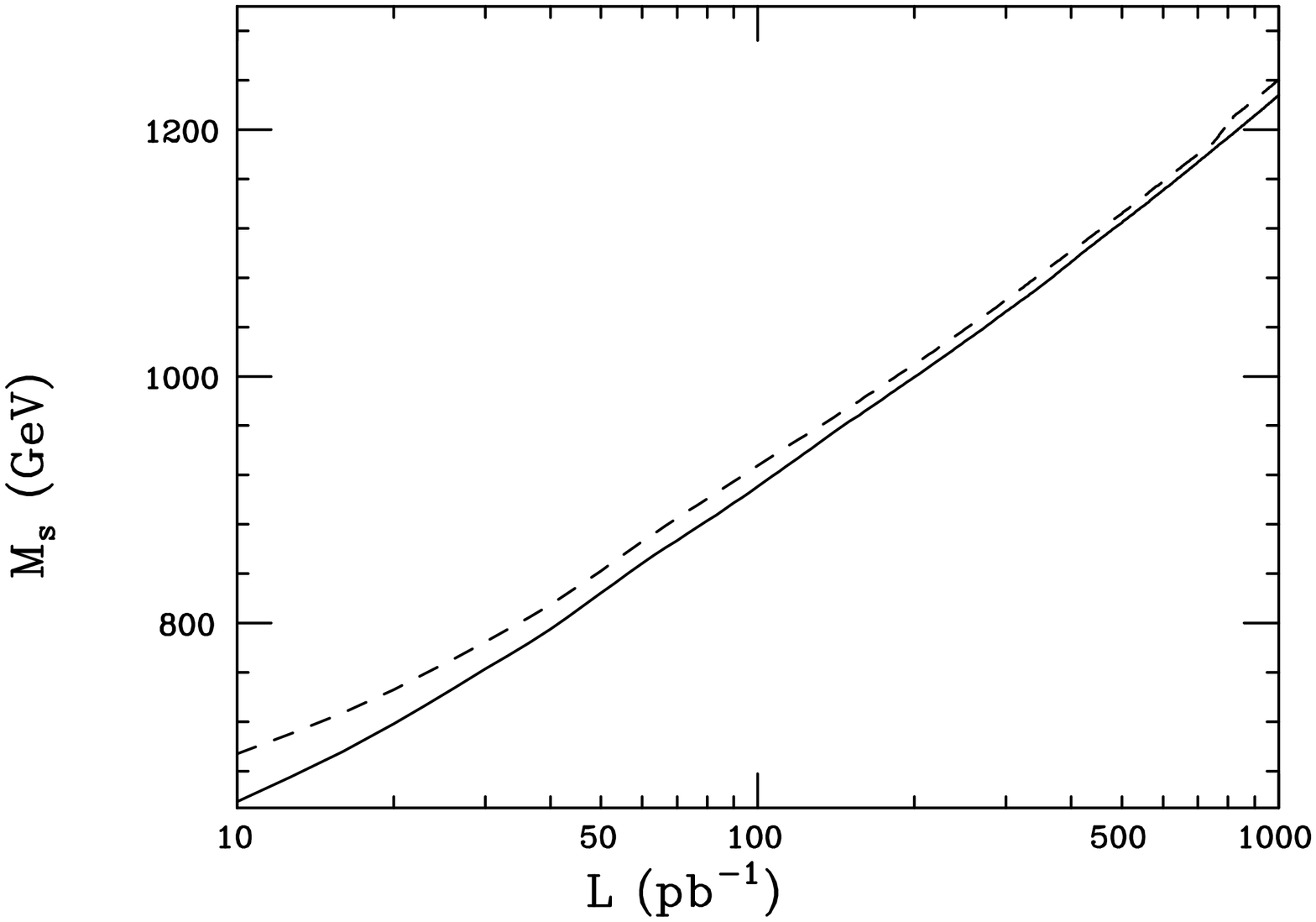,height=8cm,width=9cm,angle=-90}}
\vspace*{-1cm}
\caption[*]{$95\%$ CL lower bound on the value of $M_s$ obtainable at HERA as 
a function of the integrated luminosity per charge/polarization state for 
$\lambda=\pm 1$.}
\end{figure}
\nopagebreak[4]

While it is anticipated that the 
initial laser polarization will be near $100\%$, \ie, $|P_l|=1$, the electron 
beam polarization is expected to be be near $90\%$, \ie, $|P_e|=0.9$. We  
assume these values in the analysis that follows. With two photon `beams' 
and the choices $P_l=\pm 1$ and $P_e=\pm 0.9$ to be made for each beam it 
would appear that 16 distinct polarization-dependent cross sections need to be 
examined. However, we find that there only six physically distinct initial 
polarization combinations. In what follows we will label these 
possibilities by the corresponding signs of the electron and laser 
polarizations as $(P_{e1},P_{l1},P_{e2},P_{l2})$. (Unpolarized cross sections 
can be obtained by averaging.)
Clearly some of these polarization 
combinations will be more sensitive to the effects of K-K towers of gravitons 
than will others.

\nopagebreak[4]
\vspace*{-0.5cm}
\nn
\begin{figure}[htbp]
\centerline{
\psfig{figure=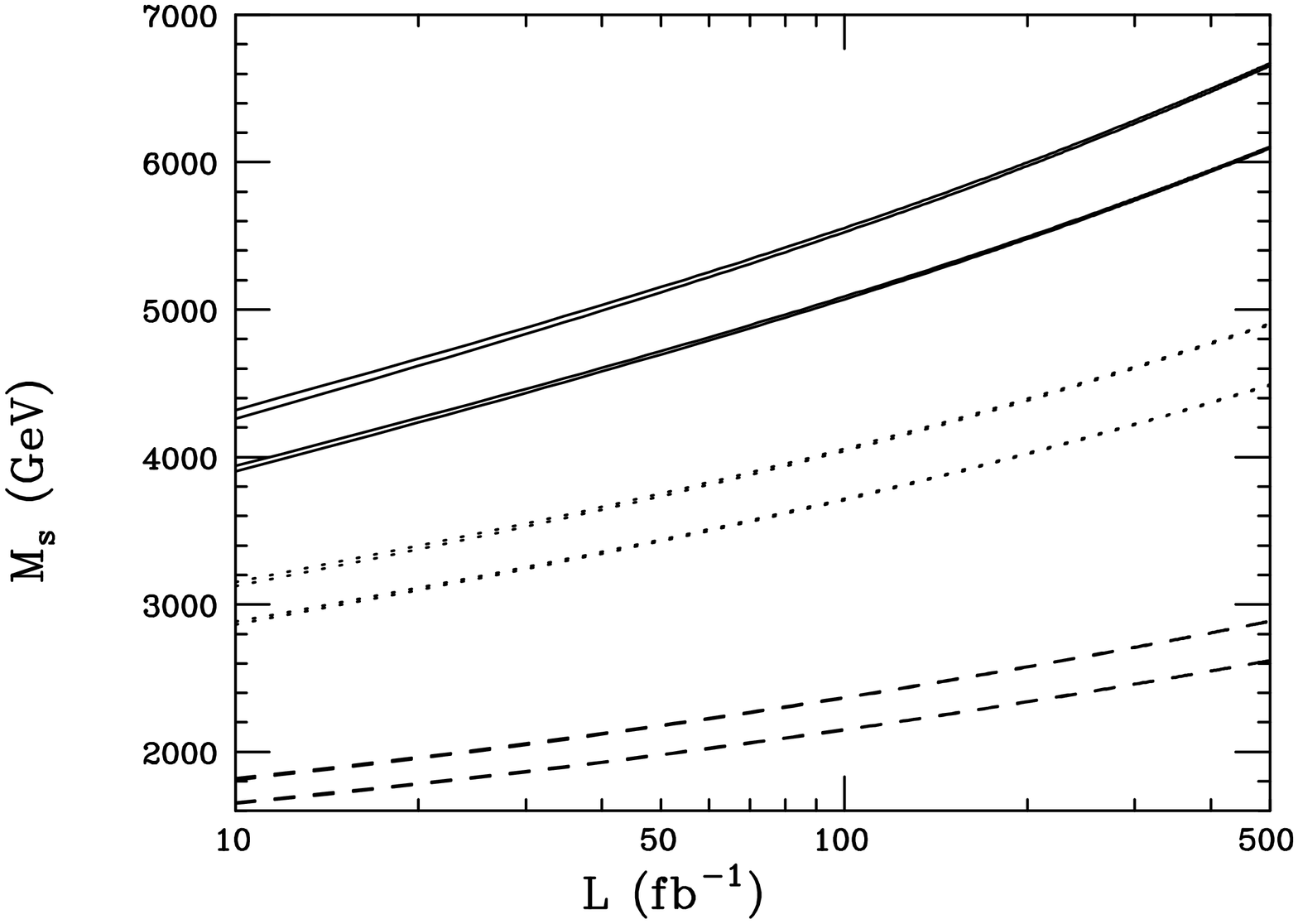,height=8.0cm,width=9cm,angle=-90}}
\vspace*{-0.6cm}
\caption{Search reaches for the processes $\gamma \gamma \to f\bar f$, 
with $f$ being the $c,t$ and $b$ quarks together with $e$, $\mu$ and 
$\tau$(lowest curve of a given type), and for lepton pairs, top, plus 
light quark jets(upper pair of 
curves) as a function of the total $\gamma \gamma$ integrated luminosity. At a 
500(1000, 1500) GeV $e^+e^-$ collider the result is given by 
the dashed(dotted, solid) curve and in the former case is 
essentially independent of the choice $\lambda=\pm$ 1.}
\end{figure}
\nopagebreak[4]

The first set of processes to examine is $\gamma \gamma \to f\bar f,gg$ which 
have so far only fully 
been examined in the case of unpolarized beams. Note that since there are 
identical particles in the initial state no $A_{FB}$ can be formed in this 
case. 
The analysis here is 
similar to that for combined channel $e^+e^-\to f\bar f$ study in that all 
accessible final states are included in the fit; the results are shown in 
Fig.12. The obtainable reach is found to be $M_s \simeq 4\sqrt s_{e^+e^-}$, 
somewhat less than in $e^+e^-$ collisions mostly due to the lower effective 
center of mass energy of the colliding photons and the lack of polarization 
information. One would naively expect that the reach for $M_s$ would increase 
somewhat if the possibility of beam polarization were included in the 
analysis.

\nopagebreak[4]
\vspace*{-0.5cm}
\nn
\begin{figure}[htbp]
\centerline{
\psfig{figure=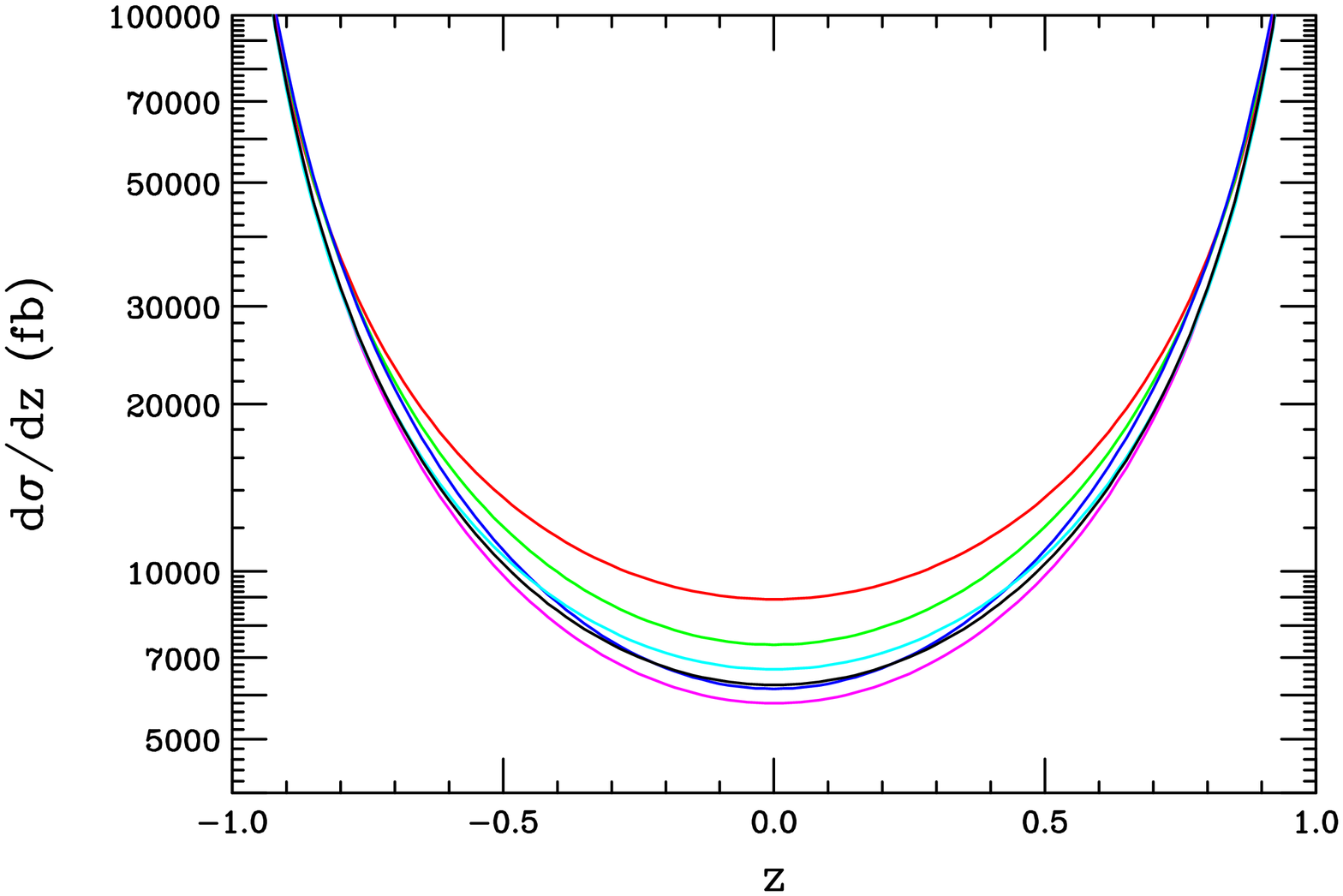,height=5.9cm,width=8cm,angle=90}}
\vspace*{0.9cm}
\centerline{
\psfig{figure=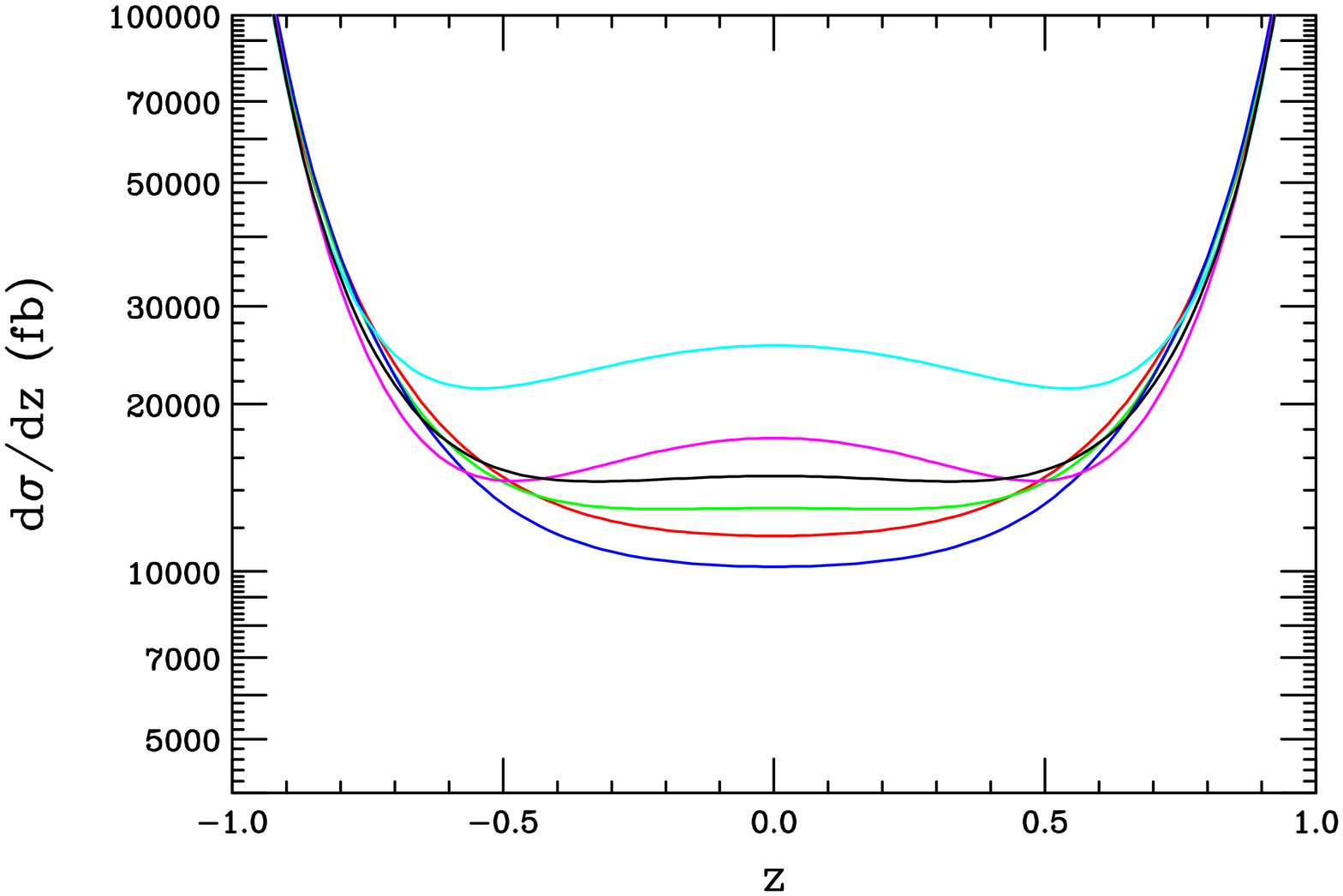,height=5.9cm,width=8cm,angle=90}}
\vspace*{0.8cm}
\caption{Differential cross section for $\gamma\gamma \to W^+W^-$ at a 1 TeV 
$e^+e^-$ collider for (Top)the SM and with $M_s=2.5$ TeV with 
(Bottom)$\lambda=1$. 
The $\lambda=-1$ results are nearly identical. 
In (Top) from top to bottom in the center of the figure 
the helicities are $(++++)$, 
$(+++-)$, $(-++-)$, $(++--)$, $(+---)$, and $(+-+-)$; in (Bottom) they are  
$(-++-)$, $(+-+-)$, $(+++-)$, $(+---)$, $(++++)$, and $(++--)$}
\end{figure}
\nopagebreak[4]

The $W^+W^-$ final state offers many observables and very high statistics 
with which to probe the KK contributions. 
The differential cross sections are shown in Fig.13 for the SM as 
well as when the K-K tower is turned for all 
six initial helicity combinations. Note that in the SM there is no 
dramatically strong sensitivity to the initial state lepton and laser 
polarizations and all of the curves have roughly the same shape. 
When the graviton tower contributions are included there are several 
effects. First, all 
of differential distributions become somewhat more shallow at large 
scattering angles but there is little change in the forward and backward 
directions due to the dominance of the SM poles. Second, there is now a 
clear and distinct sensitivity to the initial state 
polarization selections. In some cases, particularly for the $(-++-)$ and 
$(+-+-)$ helicity choices, the differential cross section increases 
significantly for angles near $90^o$ taking on an m-like shape. This shape is, 
in fact, symptomatic of the spin-2 nature of the K-K graviton tower exchange 
since a spin-0 exchange leads only to a flattened distribution. 
Given the very large statistics available with a typical integrated luminosity 
of 100 $fb^{-1}$, it is clear that the $\gamma \gamma \to W^+W^-$ 
differential cross 
section is quite sensitive to $M_s$ especially for the two initial state 
helicities specified above. 

\nopagebreak[4]
\vspace*{-0.5cm}
\nn
\begin{figure}[htbp]
\centerline{
\psfig{figure=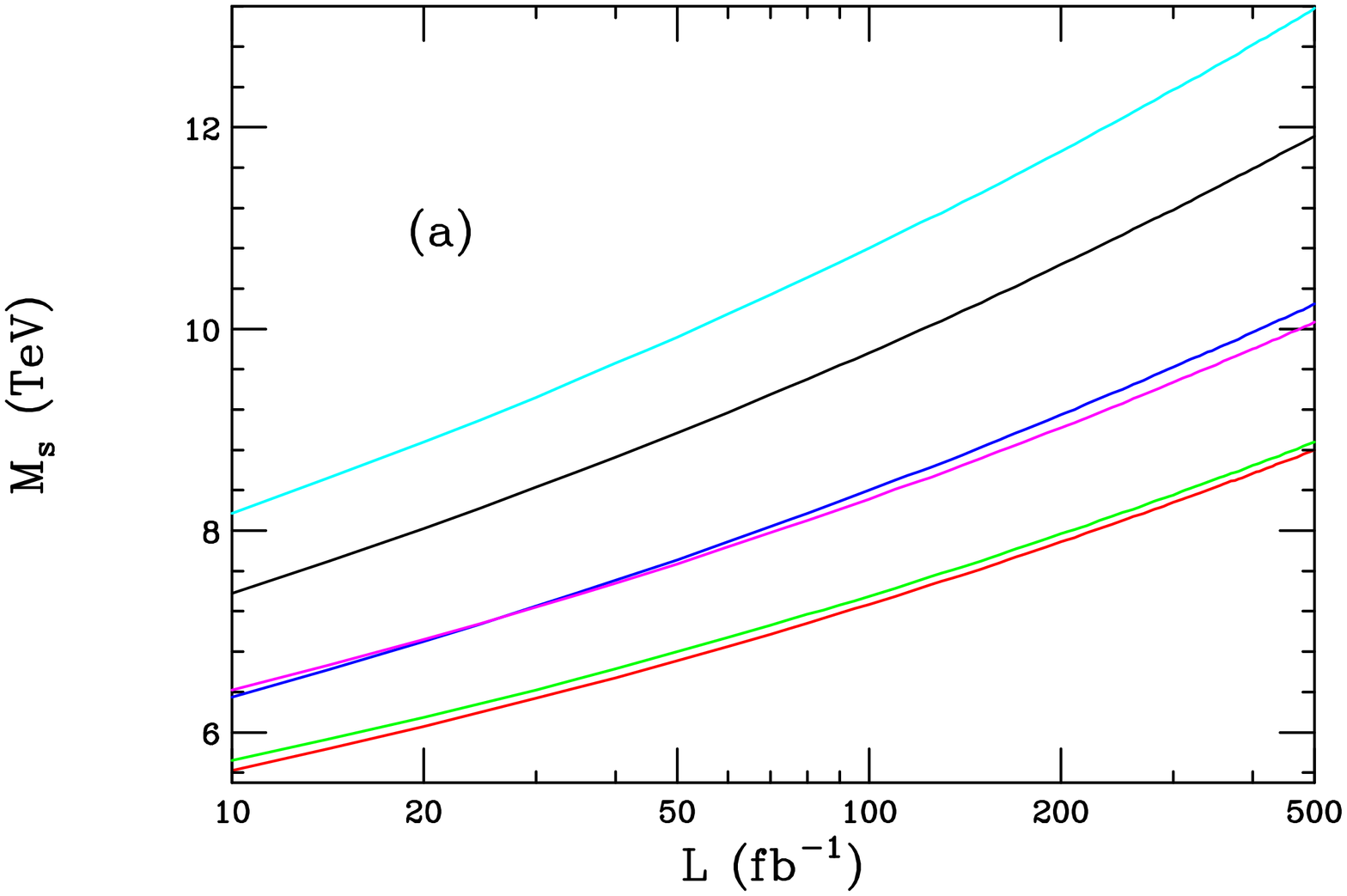,height=5.9cm,width=8cm,angle=90}}
\vspace*{0.9cm}
\centerline{
\psfig{figure=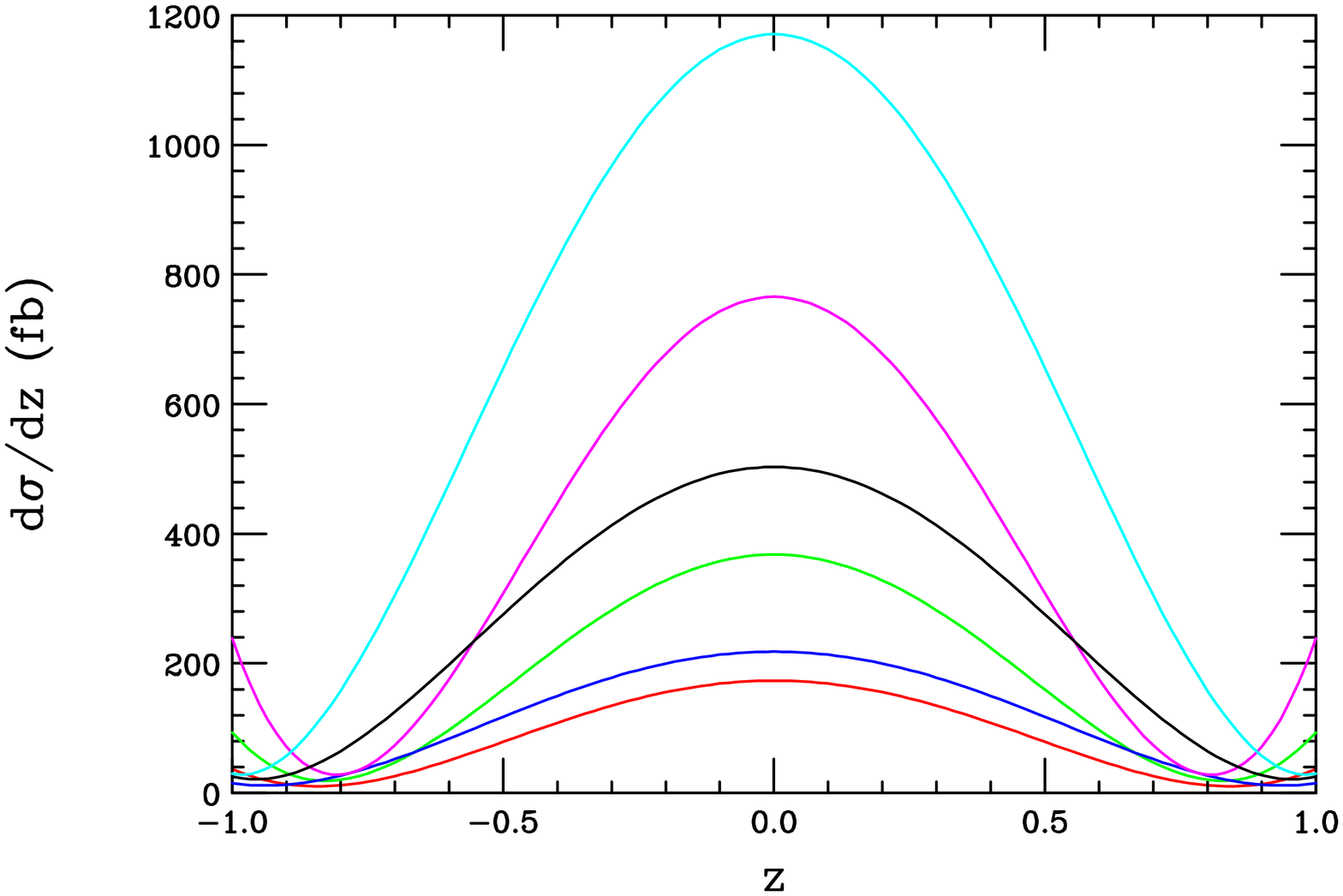,height=5.9cm,width=8cm,angle=90}}
\vspace*{0.8cm}
\caption{(Top)$M_s$ discovery reach from the 
process $\gamma \gamma \to W^+W^-$ at a 1 TeV $e^+e^-$ collider 
as a function of the integrated luminosity 
for the different initial state polarizations. From top to bottom on the 
right hand side of the figure the 
polarizations are $(-++-)$, $(+---)$, $(++--)$, $(+-+-)$, $(+---)$,  
and $(++++)$.
(Bottom)Differential cross section for $\gamma \gamma \to ZZ$ at a 1 TeV 
$e^+e^-$ collider due to the exchange 
of a K-K tower of gravitons assuming $M_s=3$ 
TeV. From top to bottom in the center of the figure the initial state 
helicities are $(-++-)$, $(+-+-)$, $(+---)$, $(+++-)$, $(++--)$, $(++++)$.}
\end{figure}
\nopagebreak[4]

In the SM, the final state 
$W$'s are dominantly transversely polarized. Due to the nature of the spin-2 
graviton exchange, the K-K tower leads to a final state where both $W$'s are 
completely 
longitudinally polarized. Thus we might expect that a measurement of the 
$W$ polarization will probe $M_s$. 
By combining a fit to the total cross sections and angular distributions as 
well as the $LL$ and $LT+TL$ helicity fractions for various initial state 
polarization choices we are able to discern the discovery as well as the 
$95\%$ CL exclusion reaches for $M_s$ as shown in Fig.14. Here we see a reach 
of $M-s\sim 11\sqrt s$, which is the larger than that obtained from all 
other processes examined so far{\cite {pheno2}}. 

The process $\gamma \gamma \to ZZ$ does not occur at the tree level in the SM 
or MSSM. This would naively seem to imply 
that this channel is particularly suitable for looking for new physics effects 
since the SM and MSSM rates will be so small due to the loop suppression; 
unfortunately this is not the case. 
In the case of the ADD scenario the tree level K-K graviton tower contribution 
is now also present. Neglecting the loop-order SM 
contributions for the moment we obtain the resulting polarization-dependent 
differential cross sections shown in Fig.14. The rather small cross section 
found here leads to a rather poor sensitivity it $M_s$ of order $4-5\sqrt s$.

\section{Summary}

In this paper we have surveyed some of the processes which can be used to 
probe for the exchange of KK towers of gravitons in the ADD model. 
Table I from {\cite {jlh}} gives an overall summary of the indirect search 
reaches for scale $M_s$.

\begin{table}
\centering
\begin{tabular}{|c|c|c|} \hline\hline
Reaction & LEP II (2 fb$^{-1}$) & LC (100 fb$^{-1}$) \\ \hline
$\epem\to f\bar f$ & 1.15 & 6.5$\sqrt s$ \\
$\epem\to\epem$ & 1.0 & 6.2$\sqrt s$ \\
$e^-e^-\to e^-e^-$ &  & 6.0$\sqrt s$ \\
$\epem\to\gamma\gamma$ & 1.4 &  3.2$\sqrt s$ \\
$\epem\to WW/ZZ$ & 0.9 & 5.5$\sqrt s$ \\ \hline
        & Tevatron (2 fb$^{-1}$) & LHC (100 fb$^{-1}$) \\ \hline
$p^(\bar p^)\to \ell^+\ell^-$ & 1.4 & 5.3 \\
$p^(\bar p^)\to t\bar t$ & 1.0 & 6.0 \\
$p^(\bar p^)\to jj$ & 1.0 & \\
$p^(\bar p^)\to WW$ & 0.8 & \\
$p^(\bar p^)\to \gamma\gamma$ & 1.4 & 5.4 \\ \hline
       & HERA (250 pb$^{-1}$) & \\ \hline
$ep\to e+$ jet & 1.0  \\ \hline
       & $\gamma\gamma$ Collider (100 fb$^{-1}$) & \\ \hline
$\gamma\gamma\to \ell^+\ell^-/t\bar t/jj$ & 4$\sqrt s$ & \\
$\gamma\gamma\to\gamma\gamma/ZZ$ & $(4-5)\sqrt s$ & \\
$\gamma\gamma\to WW$ & 11$\sqrt s$ & \\ \hline\hline
\end{tabular}
\caption{$M_s$ search limits in TeV for a number of various processes.}
\label{exch}
\end{table}

\section*{Acknowledgments}
The author would like to thank J.L. Hewett for her help in preparing this 
review.
%
\def\MPL #1 #2 #3 {Mod. Phys. Lett. {\bf#1},\ #2 (#3)}
\def\NPB #1 #2 #3 {Nucl. Phys. {\bf#1},\ #2 (#3)}
\def\PLB #1 #2 #3 {Phys. Lett. {\bf#1},\ #2 (#3)}
\def\PR #1 #2 #3 {Phys. Rep. {\bf#1},\ #2 (#3)}
\def\PRD #1 #2 #3 {Phys. Rev. {\bf#1},\ #2 (#3)}
\def\PRL #1 #2 #3 {Phys. Rev. Lett. {\bf#1},\ #2 (#3)}
\def\RMP #1 #2 #3 {Rev. Mod. Phys. {\bf#1},\ #2 (#3)}
\def\NIM #1 #2 #3 {Nuc. Inst. Meth. {\bf#1},\ #2 (#3)}
\def\ZPC #1 #2 #3 {Z. Phys. {\bf#1},\ #2 (#3)}
\def\EJPC #1 #2 #3 {E. Phys. J. {\bf#1},\ #2 (#3)}
\def\IJMP #1 #2 #3 {Int. J. Mod. Phys. {\bf#1},\ #2 (#3)}

\end{document}